%% file: snoopy.tex
\documentclass[acmlarge]{acmart}

\usepackage{booktabs} 
\usepackage[ruled]{algorithm2e} 
\usepackage{graphicx}
\usepackage{subcaption}
\usepackage{tabularx}
\usepackage{epsfig}  
\usepackage{xcolor}
\usepackage{multirow}
\usepackage{amssymb}
\usepackage{appendix}
\usepackage{xcolor}
\usepackage{pdfpages}
\usepackage{pifont}
\newcommand{\xmark}{\ding{55}}%

\newcommand{\chris}[1]{{\color{black}{#1}}}
\newcommand{\andrew}[1]{{\color{black}{#1}}}
\newcommand{\hongkai}[1]{{\color{black}{#1}}}

\newcommand{\eat}[1]{}

\SetAlFnt{\small}
\SetAlCapFnt{\small}
\SetAlCapNameFnt{\small}
\SetAlCapHSkip{0pt}
\IncMargin{-\parindent}

\acmJournal{IMWUT}
\acmVolume{1}
\acmNumber{4}
\acmArticle{152}
\acmYear{2017}
\acmMonth{12}
\acmArticleSeq{11}
\acmDOI{10.1145/3161196}
\acmPrice{15.00}

\setcopyright{acmcopyright}

\begin{document}
\title{Snoopy: Sniffing Your Smartwatch Passwords via Deep Sequence Learning} 
\author{Chris Xiaoxuan Lu}
\authornote{Email: \href{mailto:luxiaoxuan555@hotmail.com}{luxiaoxuan555@hotmail.com} }
\affiliation{%
  \department{Department of Computer Science}
  \institution{University of Oxford}
  \city{Oxford}
  \postcode{OX1 3QD}
  \country{UK}
}
\author{Bowen Du}
\author{Hongkai Wen}
\authornote{Hongkai Wen is the corresponding author.}
\affiliation{%
  \institution{University of Warwick}
  \department{Department of Computer Science}
  \city{Coventry}
  \postcode{CV4 7AL}
  \country{USA}
}
\author{Sen Wang}
\affiliation{%
  \department{Department of Computer Science}
  \institution{Heriot-Watt University}
  \city{Oxford}
  \postcode{OX1 3QD}
  \country{UK}
}
\author{Andrew Markham}
\author{Ivan Martinovic}
\affiliation{%
  \department{Department of Computer Science}
  \institution{University of Oxford}
  \city{Oxford}
  \postcode{OX1 3QD}
  \country{UK}
}
\author{Yiran Shen}
\affiliation{%
  \department{College of Computer Science and Technology}
  \institution{Harbin Engineering University}
  \city{Harbin}
  \postcode{150001}
  \country{China}
}

\author{Niki Trigoni}
\affiliation{%
  \department{Department of Computer Science}
  \institution{University of Oxford}
  \city{Oxford}
  \postcode{OX1 3QD}
  \country{UK}
}

\input{section/abstract}

%
%


%
%



\thanks{}

\maketitle
%
\renewcommand{\shortauthors}{C. X. Lu et al.}

\input{section/intro}

\input{section/bkg}

\input{section/overview}
\input{section/pwd_ext}

\input{section/pwd_inf}
\input{section/eval}

\input{section/us}

\input{section/disc}
\input{section/related_work}

\input{section/conclution}

\bibliographystyle{ACM-Reference-Format}
\bibliography{acmart}

\end{document}

%% file: section/abstract.tex
\begin{abstract}
Demand for smartwatches has taken off in recent years with new models which can run independently from smartphones and provide more useful features, becoming first-class mobile platforms. One can access online banking or even make payments on a smartwatch without a paired phone. This makes smartwatches more attractive and vulnerable to malicious attacks, which to date have been largely overlooked. In this paper, we demonstrate Snoopy, a password extraction and inference system which is able to accurately infer passwords entered on Android/Apple watches within 20 attempts, just by eavesdropping on motion sensors. Snoopy uses a uniform framework to extract the segments of motion data when passwords are entered, and uses novel deep neural networks to infer the actual passwords. We evaluate the proposed Snoopy system in the real-world with data from 362 participants and show that our system offers a $\sim3$-fold improvement in the accuracy of inferring passwords compared to the state-of-the-art, without consuming excessive energy or computational resources. We also show that Snoopy is very resilient to user and device heterogeneity: it can be trained on crowd-sourced motion data (e.g. via Amazon Mechanical Turk), and then used to attack passwords from a new user, even if they are wearing a different model. 

This paper shows that, in the wrong hands, Snoopy can potentially cause serious leaks of sensitive information. By raising awareness, we invite the community and manufacturers to revisit the risks of continuous motion sensing on smart wearable devices. 
\end{abstract}

%% file: section/intro.tex
\section{Introduction} 
\label{sec:introduction}
\begin{figure*}[!t]
\centering
\includegraphics[width=0.8\columnwidth]{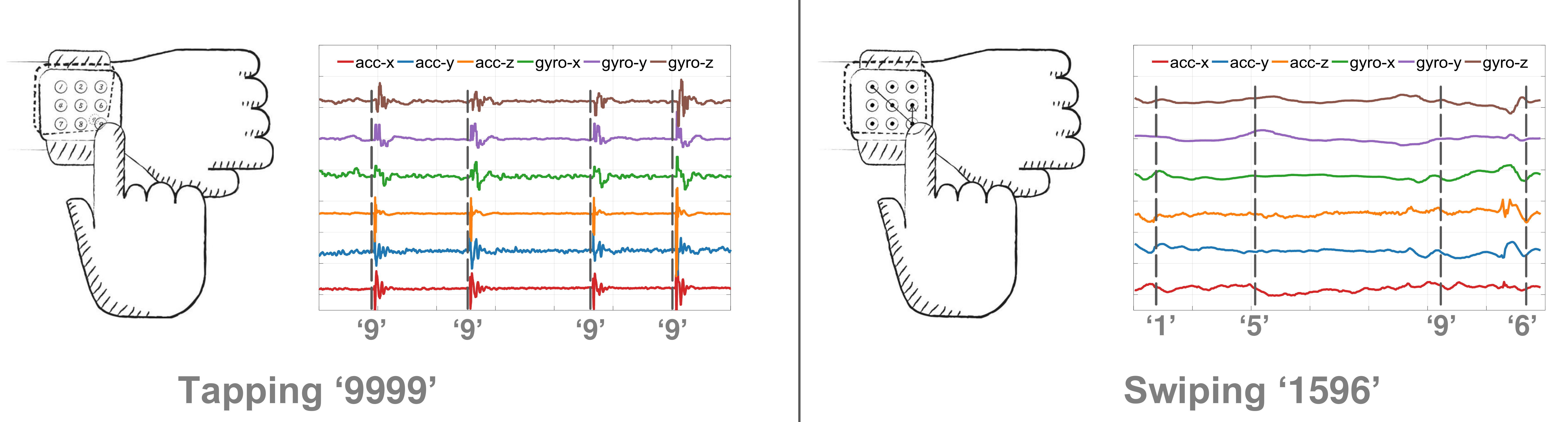}
\caption{An example of motion sensor data changes induced by \eat{tapping (Left) and} swiping \eat{(Right)} a pattern-lock on a smartwatch. Tapping or swiping passwords does not follow uniform motion and is very challenging to distinguish individual digits, let alone reveal the entire code. }
\label{fig:illus}
\end{figure*}

Smartwatches are becoming increasingly ubiquitous: it is expected that the global smartwatch market has a potential to reach \$32.9 billion by 2020 \cite{watch2020}. They are now deeply embedded in our daily lives, and over time can accumulate a variety of sensitive and important information such as emails, contacts and payment details. Due to their current role as an extension to the smartphones, the security and privacy of smartwatches have been overlooked, and instead delegated to the paired phones. However, driven by the major players such as Google and Apple, smartwatches are becoming more independent and can act as first class citizens in the mobile ecosystem: they are no long just secondary displays, but are able to offer all basic functionalities without the presence of smartphones. For instance, it is already possible to pay via a smartwatch without even needing to carry a smartphone~\cite{watch_pay,alipay_app,pay_setup}, and many recent apps on smartwatches such as fitness tracking, well-being monitoring, and messaging (email/text) apps can work independently of phone usage.

These increased functionalities make smartwatches more useful, but also attract malicious attacks which traditionally target smartphone class devices only. \andrew{Smartwatches are typically secured using a 4 digit PIN or a pattern lock, e.g., Android Pattern Locks (APLs). These are used not only to unlock the phone, but also to authenticate the payments.} In practice, the consequences of such an attack can be more serious than just security breach of the smartwatch screen lock~\cite{stobert2014password}: as shown in our user study (discussed in Sec.~\ref{sec:user_study}), over 80\% of 745 anonymous participants have a frequent habit of reusing the same passwords across services e.g. PayPal, card payments e.g. ATM PIN codes or even physical security e.g. home alarm systems. Therefore compromising a smartwatch password could lead to a series of cyber and physical attacks. 

There has been a solid body of work on the similar problem of attacking passwords on smartphones, including analysing oily residues on the screen~\cite{aviv2010smudge}, video footage~\cite{ye2017cracking}, radio signal perturbations~\cite{li2016csi} and motion sensor data~\cite{cai2012practicality}. In particular, attacking passwords by eavesdropping motion data is popular, since motion sensors are commonly sampled by a wide variety of applications, e.g. those designed for positioning, fitness tracking and activity recognition. In addition, giving access to on-board motion sensors appears to be innocuous, and many users (76\% according to our user study) would grant instantly. \andrew{Motion sensors leak information about the location of events such as taps through small changes in orientation and impacts. Existing techniques for cracking PINs on smartphones typically segment digits by extracting tap events and then use the extracted features to tell which digit has been pressed. As such, existing techniques rely on significantly handcrafted features and ad hoc approaches for digit segmentation. } 

\begin{figure*}[t]
\centering
\begin{subfigure}[b]{0.4\textwidth}\centering
\includegraphics[width=\columnwidth]{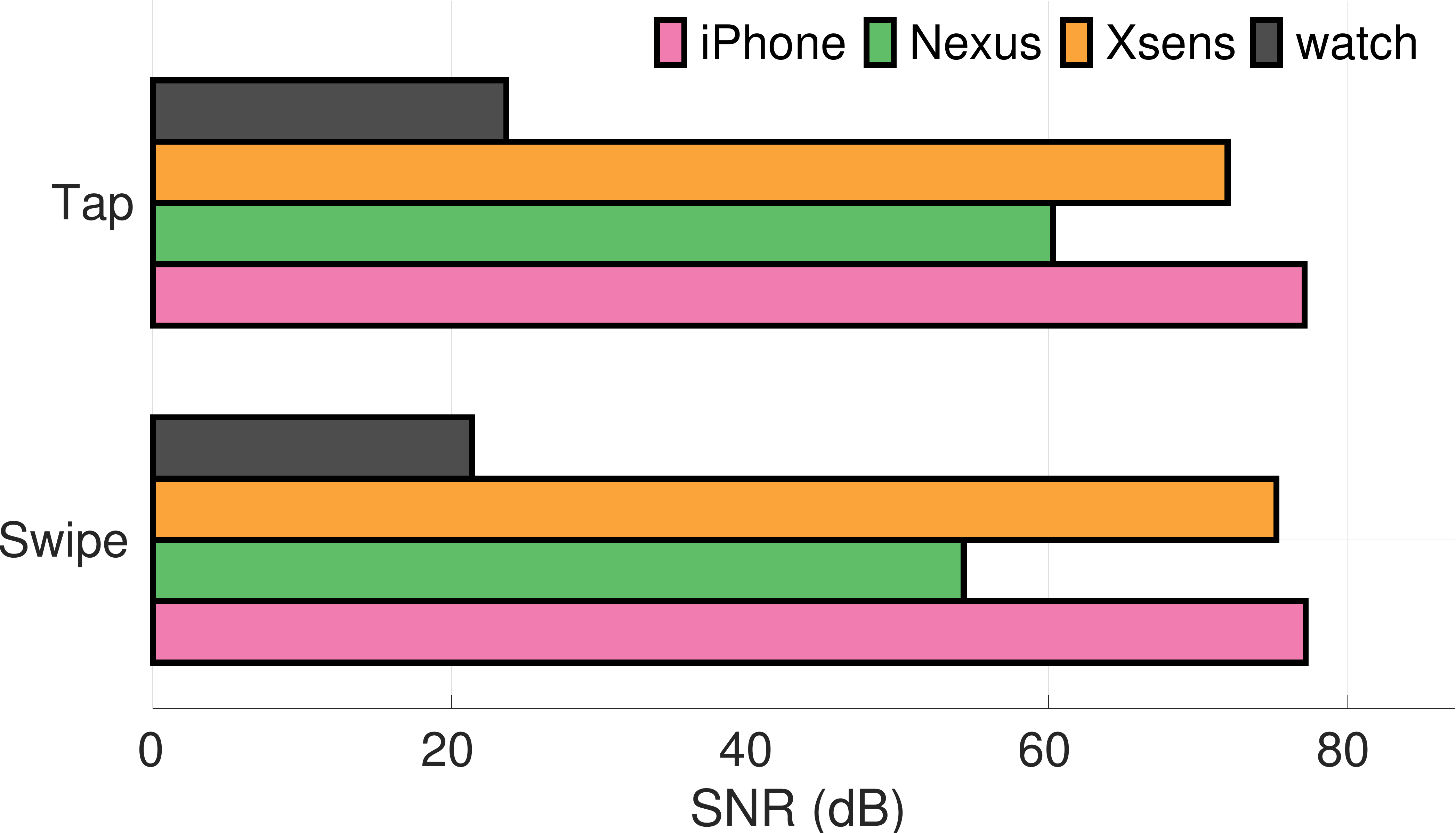}  
\end{subfigure}%
\hspace{5em}
\centering
\begin{subfigure}[b]{0.4\textwidth}\centering
\includegraphics[width=\columnwidth]{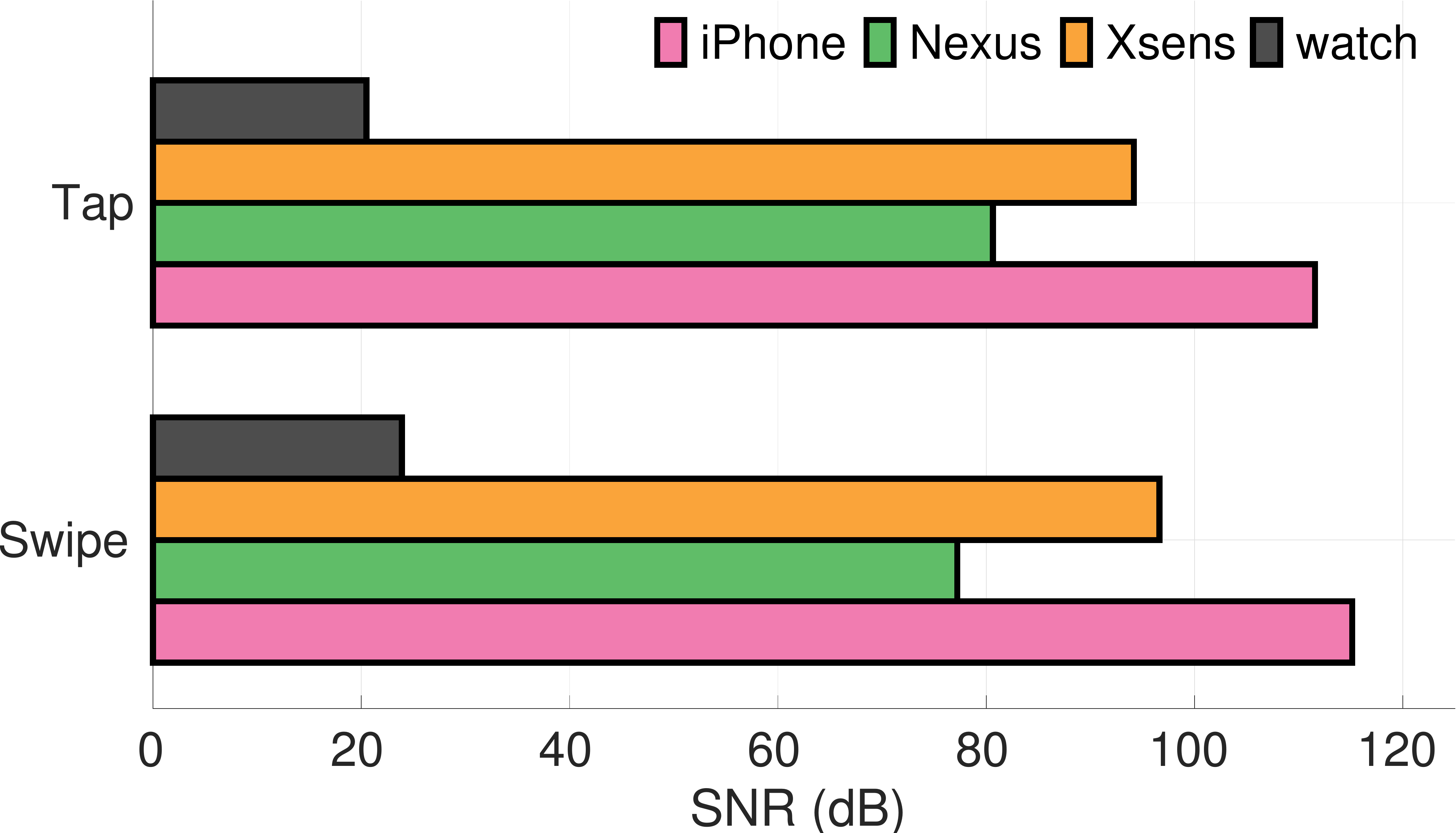} 
\end{subfigure}%
\caption{Signal-to-noise ratio (SNR) of motion sensors on smartphones, high-end IMUs and smartwatches. Left: Accelerometers; Right: Gyroscopes}
\label{fig:snr}
\end{figure*}

\andrew{Smartwatches with their smaller form factor compound the password classification problem, making it far more challenging than the smartphone case.}
Fig.~\ref{fig:illus} shows examples where the motion sensor data changes as the user \eat{taps a PIN or} swipes an APL on a smartwatch. As can be seen, the motion  induced by password entries is small and not easily segmented. Fig.~\ref{fig:snr} highlights this by considering the signal-to-noise (SNR) ratio of motion sensors on different devices. We see that motion signals on smartwatches are far noisier, and can be 20-40dB worse than that of smartphones or high-end IMUs. In the presence of such low SNR, existing techniques designed for smartphones~\cite{cai2012practicality} typically fail to work. \eat{The first reason is that they cannot accurately segment out pertinent part of the signal that corresponds to the password entry on smartwatches. More importantly, even if the password signal is correctly extracted,} \andrew{This is due to the reliance on hand-engineered features, which are not robust to variability across users and devices, particularly given the much weaker and noisier motion signals on smartwatches.} 

\eat{the inference of the actual password is still problematic since they use hand engineered features, which are not robust to the variability in users and devices,}

\andrew{Although there are a number of papers which look at cracking PINs on smartphones, to date, only one paper has considered the issue of revealing APLs. This is because APLs can have an arbitrary length and hence have significantly more possible combinations e.g. 60 passwords vs 10k for a 4 digit PIN. In this paper, we provide a \emph{universal} data driven technique for inferring both APLs and PINs on smartwatches. This requires no handcrafted feature extraction or digit segmentation and is able to generalize well to the problem of arbitrary length APLs, even when faced with the extremely low SNRs found on smartwatches. Our novel deep learning approach, based on recurrent neural networks (RNNs), exhibits a 3-4 fold increase in accuracy compared with the state-of-the-art. We propose two different architectures. The first exploits the skewed distribution of passwords to perform entire passwords inference. This technique shows superior results on popular passwords. The second is capable of digit level inference i.e. it can generalize to any password that may or may not be present in a training database. Although we have only considered the challenges of smartwatches in this paper, this technique would easily work on the high SNR signals found on smartphones as well. In addition, we propose an adaptive motion sensing technique to  detect the password in low sampling rate and it only sends motion data of the candidate password events to the server, minimizing battery and bandwidth usage to mimic a Trojan fitness app more readily. In summary, the contributions of this paper are:}

\eat{In this paper, we address both the challenges of password extraction and password inference. In terms of extraction, we provide a uniform pipeline of classification and smoothing techniques that works for both tapped PINs and swiped APLs.}

\eat{With respect to inference, we propose a deep learning approach, based on recurrent neural networks (RNNs), which exhibits 3-4 fold increase in accuracy compared to existing methods. We have designed different RNN architectures to handle the unique challenges associated with tapped PINs vs. swiped APLs. Although we have only considered the challenges of smartwatches in this paper, this technique would easily work on the high SNR signals found on smartphones as well. Finally, our system attempts to minimise battery and bandwidth usage to make the attack realistic. The assumption is that greedy use of resources could help uncover the illegitimate eavesdropping of passwords and compromise the attack. We therefore show that such attacks are not only possible, but can also be designed to be imperceptible by unsuspicious users. To summarise, our paper makes the following contributions:}

\begin{itemize}

\item  We present Snoopy, the first system that demonstrates the feasibility of intercepting password information entered on smartwatches by sensing resulting motion data. 

\item \andrew{Snoopy is also the first approach that can infer universal APLs, a significantly more difficult problem than PINs, due to the challenges of digit segmentation. }


\item  \andrew{We propose a \emph{universal} password inference mechanism based on deep recurrent neural networks that is able to attack PINs and APLs. We present two variants, one which cracks popular passwords and another which infers arbitrary passwords. Our system does not require any handcrafted features, only a crowdsourced training dataset.}

\eat{novel password inference method based on deep recurrent neural networks for both tapped PINs and swiped APL, which can achieve high inference accuracy across users and smartwatch types.}

\item \chris{We have conducted a user study and collected over 1,000 answered questionnaires, which shows that the affected population of smartwatch users is nonnegligible and the majority of users are not aware of the potential password leak on smartwatches via motion data and its consequences.}

\item We have extensively evaluated the proposed Snoopy system, using data from over $360$ distinct participants and $>60$K password entries on both Android and Apple devices. Our results show that the Snoopy achieves a 3-4 fold improvement in correctly inferred passwords compared to competing techniques.

\end{itemize}

The remainder of the paper is organised as follows. Sec.~\ref{sec:bkg} covers the technical background, while Sec.~\ref{sec:sys_overview} presents the overview of the proposed Snoopy system. Sec.~\ref{sec:pwd_ext} proposes a uniform approach to extracting the relevant part of the motion signal that corresponds to entering PINs and APLs. Sec.~\ref{sec:pwd_inf} proposes two models of inferring the actual contents of passwords using deep neural networks. Sec.~\ref{sec:evaluation} evaluates the proposed extraction and inference approaches and compares them with competing state-of-the-art methods. Sec.~\ref{sec:user_study} reports the results of our user study, while Sec.~\ref{sec:discussion} discusses the energy/accuracy tradeoff of Snoopy and possible countermeasures. Finally, Sec.~\ref{sec:related_works} presents an overview of related work, and Sec.~\ref{sec:conclusion} concludes the paper and points to directions for future work.

%% file: section/bkg.tex

\section{Technical Background} 
\label{sec:bkg}

\subsection{Tapped vs. Swiped Passwords}
There are two predominant types of password input mechanism on smartwatches (also on smartphones): tapped and swiped passwords~\cite{von2013patterns}. 

For iOS platforms, the default password type is four digit PIN, where the users \emph{tap} their passwords on the screen when prompted. A four digit PIN has $10,000$ possible combinations. It is possible to use longer passwords, but in this paper, we only consider the  four digit PIN.

For the Android platform, users have the option to use a PIN or a graphical pattern lock (also known as APL), where the users \emph{swipe} a pattern over a three-by-three matrix of dots (see Fig.\ref{fig:illus} for an example). Unlike the numerical passwords where the users can choose freely from ten possible digits at each tap, the smartwatch operating systems typically have certain constraints over the trajectories of the swiped patterns. For instance as shown in Fig.\ref{fig:illus}, starting from the top left dot, it is only possible to swipe towards four reachable neighbours: the immediate right and the three dots in the second row. Therefore, the size of the search space for swiped passwords is restricted to a maximum of $389,112$~\cite{aviv2010smudge}. 

In practice, for both PINs and APLs, if one fails to input the correct password three times the smartwatch will prevent any further attempts for a few minutes. When the number of failed attempts reaches a threshold, typically ten, the smartwatch can enter `lost' mode, e.g. erase all data.

\subsection{Motion Induced by Password Input}
Intuitively, entering both tapped and swiped passwords will induce forces and orientation changes on the smartwatch \cite{turner2017text}. Since human skin has a certain level of elasticity, tapping on the smartwatch screen will cause minor displacement at the contact point along the vertical direction, i.e. the watch body will rotate for a small angle. Tapping causes an underdamped impulsive wave to develop, which causes small oscillations, shown in Fig.~\ref{fig:illus}. On the other hand, when swiping passwords, the pressing and friction force between the user's finger and touch screen will ``drag'' the smartwatch to move along both vertical and horizontal directions. This gives rise to small slip-pulse waves which have a longer duration than impulsive taps, as shown in Fig.~\ref{fig:illus}.

In practice, induced motion can be picked up by the Inertial Measurement Units (IMUs) embedded on most of the commercial smartwatches. IMU sensors have been widely used in many mobile sensing scenarios, since they are able to capture displacement and rotation of the devices in 3-D space, and become increasingly cheap and power efficient. Concretely in this work we consider both accelerometers and gyroscopes, which capture the linear acceleration and angular velocity (roll, pitch and yaw) with respect to the three axis. By default the IMU sensors on most smartwatches are set to be always-on, continuously sensing motion for various applications such as gesture recognition, localisation, and fitness monitoring. This can lead to involuntary information leakage, which may be leveraged by malicious parties to infer private and valuable data such as passwords. 


%% file: section/overview.tex
\section{System Overview}
\label{sec:sys_overview}

\subsection{Attack Assumptions}
We assume that the user installs Snoopy, a Trojan app that can be easily disguised as a fitness or gaming app~\cite{liu2015good}. Snoopy requires access to the motion sensors, which does not require explicit permission in Android or via CMMotionManager in iOS. Snoopy logs and periodically sends candidate extracted password events via the network. In Android this is given by the \emph{INTERNET} permission which is classed as a normal permission, not a dangerous permission. In iOS, this is done via a normal system API and thus no additional permissions are required. The amount of data that needs to be sent is also small - a 10s batch of candidate password data is only 3~kByte, so Snoopy is unlikely to trigger any network level monitors. Note that throughout the attack, Snoopy only needs to eavesdrop motion data, without having access to any other sensing modality, such as monitoring the touch screen~\cite{ye2017cracking,shukla2014beware}.

\subsection{Attack Goals}

There are two key goals of the attack. The first is to successfully harvest candidate password events for later extraction. This is because sending raw motion data without segmentation will cause heavy traffic loads and lead to the app more likely being flagged. The second is to be able to infer both tapped PINs and swiped APLs. Once comprised, an attacker can unlock the physical device, accessing all stored information. Alternatively as we observe in our user study (Sec.~\ref{sec:user_study}), many users would use the same passwords across different accounts and platforms, where compromising one password can be harmful to many services. 

\begin{figure*}[!t]
\centering
\includegraphics[width=0.98\columnwidth]{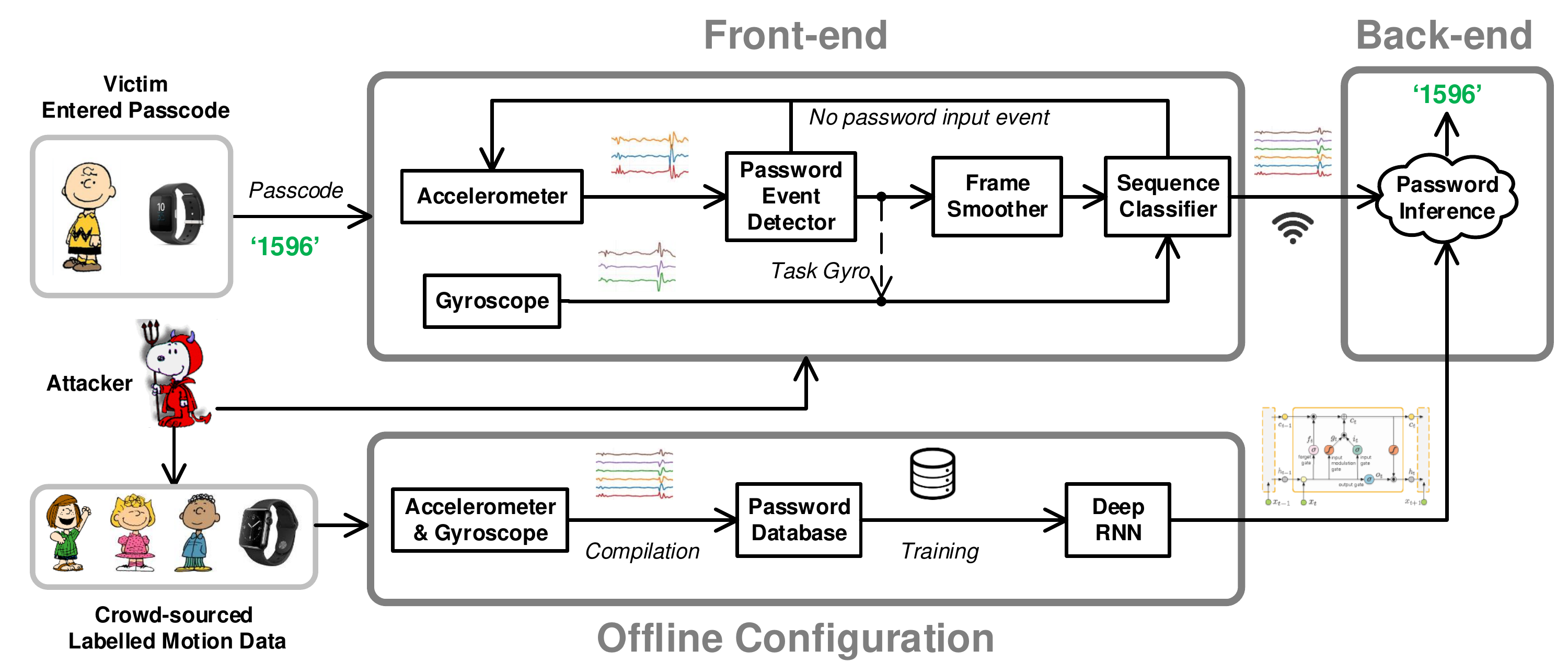}
\caption{System overview:The attacker builds a deep RNN classifier using crowd-sourced data. On a victim's smartwatch, a trojan app uses an adaptive sampling scheme to record and identify a victim's motion data. Candidate password sequences are uploaded to the server. The back-end server runs the trained deep RNN classifier to infer possible passwords. Note, training is only required by the attacker; no training is needed by the victim.}
\label{fig:sys_overview}
\end{figure*}

\subsection{System Architecture}
\label{sec:arch}
In this section we present the high-level architecture of \emph{Snoopy}, a system for inferring PINs and APLs on smartwatches. Snoopy contains a client front-end which runs locally on a victim's smartwatch which periodically sends motion data to a password inference back-end that resides on the cloud. Fig.\ref{fig:sys_overview} shows the the architecture of the proposed system. 

\noindent \textbf{Front-end Password Input Extraction: }
The front-end of Snoopy disguises itself as a harmless app, such as fitness app, and runs in the background continuously once installed. It listens to the IMU sensors and tries to detect when users are tapping or swiping passwords on their watches. To avoid being flagged as malicious by the host OS, the front-end of Snoopy uses an adaptive motion sensing strategy. It continuously samples the accelerometers at low rates to detect potential password input events. This conserves power, as accelerometers are typically one to two orders of magnitude more power efficient than gyroscopes. Once a candidate event has been detected, it enables the gyroscope and increases the sampling rate of both sensors, logging motion data until the user finishes entering passwords. Then the segment of data is smoothed and passed through a lightweight classifier, to determine retrospectively if it corresponds to a true password input event, or other user interactions such as swiping down to check notifications. In the latter case the data segment is simply discarded, while the data of true password input events is transmitted to the back-end for further analysis.

\noindent \textbf{Back-end Password Inference: }
Given extracted segments of motion data, the back-end of Snoopy aims to infer user entered passwords. Instead of relying on bespoke signal processing algorithms which require hand-crafted features and tuning, Snoopy considers an end-to-end deep learning approach, which takes the raw motion data as the input, and computes the most likely password that the users has entered. To achieve this, Snoopy extends standard deep Recurrent Neural Networks (RNNs) to capture the unique characteristics of device motion induced by tapped PINs and swiped APLs. For PINs, it uses a hierarchical RNN with two layers to filter out the motion gaps (i.e. when the user lifts her finger off the touchscreen in between two taps) before inference, while for APLs it considers a bidirectional RNN to model the long continuous motion caused by swiping patterns. Now we are in a position to present the proposed Snoopy system in more detail. 

%% file: section/pwd_ext.tex
\section{In Situ Password Extraction} 
\label{sec:pwd_ext}
From a high level point of view, Snoopy infers the users' passwords by collecting and analyzing the motion data generated on their smartwatches. Of course one can task the motion sensors continuously at high sampling rates, and stream the sensor data to the cloud for password inference. However in practice, this will incur significant cost in energy, computation and communication, where the smartwatch operating systems (Android Wear or WatchOS) can easily detect such unusual behaviour and kill Snoopy instantly. To make our attack realistic, we would like Snoopy to be as ``benign'' as possible, i.e. it should not ask for excessive battery or bandwidth use most of time, but only become active (processing/transmitting) when the users are actually inputting passwords. In Secs.~\ref{sub:adaptive_sensing} and~\ref{sub:pwd_detection}, we first explain how to adaptively task the motion sensors to detect potential password input events without incurring heavy load on the system. Then, in Sec.~\ref{sub:seq_smoothing} we discuss how to extract the precise segments of motion data corresponding to those detected candidate events, and identify if the segments are related to actual password input events.

\subsection{Adaptive Motion Sensing}
\label{sub:adaptive_sensing}
Snoopy uses the onboard accelerometers to detect potential password input events, since they are very power efficient compared with gyroscopes \cite{xiao2014lightweight}. Concretely, we consider an adaptive sensing strategy, which switches between three modes: \emph{passive listening}, \emph{password input monitoring}, and \emph{motion data extraction}, depending on different user behaviour. Most of the time Snoopy stays in the passive listening mode, where it only samples accelerometer data at low rates and runs a gesture detection algorithm. Note that in this mode, Snoopy won't necessarily incur extra load on the sensors, since in practice major smartwatch platforms have their own gesture recognition or fitness services running in the background, which already task the accelerometer continuously. When it detects a user's intention to interact with their device, via detection of a characteristic wrist movement, Snoopy transitions into the password input monitoring mode, where it increases the accelerometer sampling rate to look for potential password input events. It keeps analysing the received acceleration data, seeking to detect when the user will start entering their password. Once such an event is detected, Snoopy immediately turns into the motion data extraction mode, and samples both accelerometer and gyroscope at higher rates until it detects that the user has finished typing/swiping passwords. This segment of motion data is cached locally and passed through a classifier which decides if it corresponds to normal tapping/swiping, e.g. check email notifications, or a true password input event. In the latter case, the cached data is sent to the cloud for password inference. At the end of this process, Snoopy goes back to passive listening. In this way, Snoopy only actively processes and transmits short bursts of password related motion data, and avoids unnecessarily alerting the OS or malware monitoring frameworks.

\subsection{Password Input Event Detection}
\label{sub:pwd_detection}
As discussed above, when the users try to interact with their watches, Snoopy increases the accelerometer sampling rate and starts to check if any password is entered. Given the raw acceleration stream, Snoopy uses a sliding window of length $T$ and stride $S$ (both expressed in terms of samples) to segment the data into \emph{frames}. Each frame contains $T$ data points and the overlap between adjacent frames is $T-S$ (assuming $T\geq S$). In practice, the optimal $T$ and $S$ depend on the accelerometer sampling rate and can be learned from the data. For instance in our experiments, when the accelerometer rate is set to 40~Hz, the best $T$ and $S$ are 60 and 6 measurements respectively. Then for each frame, we would like to decide whether the user starts to input passwords within that frame. To achieve this, we first extract various features of the data frame, e.g. moments, maximum/minimum, skewness, kurtosis of individual acceleration axis, and different norms (e.g. $l_1$, $l_2$, Infinity and Frobenius norms) across all three axes. In the current Snoopy implementation we consider 41 features in total. Based on the extracted feature vector, we consider a Support Vector Machine (SVM) to label if the current frame belongs to a password input event. If so, Snoopy switches to the motion data extraction mode, which samples both the accelerometer and gyroscope at a high rate (e.g. 200Hz) and caches the data locally. This continues until it observes a sequence of consecutive frames that are not labelled as password input. In this way, Snoopy tends to save the motion data of as many potential password input events as possible. In what follows, we show how to extract the true password input event through sequence alignment and classification.

\subsection{Frame Smoothing and Password-positive Sequence Identification}
\label{sub:seq_smoothing}
Given a cached sequence of data frames, which correspond to a potential password input event, Snoopy needs to decide: a) the accurate starting and ending frames of this event, and b) if this candidate event is a true password input event or not. For the former task, we use a smoother to align the sequence of frames based on labels of nearby frames. The intuition is that labels of adjacent frames should be consistent, i.e. a chunk of frames should either belong to a password input event or not, but not have many interleaving labels. In Snoopy we consider two types of smoothers, one based on a Hidden Markov Model (HMM) to exploit the temporal correlations, and the other based on moving average (essentially majority voting). From the output of the smoother, Snoopy extracts the longest segment of frames whose labels are positive. If the segment length exceeds a minimum threshold, Snoopy considers this segment of motion data to be able to cover the potential password input event precisely. In Sec.~\ref{sec:evaluation} we will show why this smoothing process is crucial, and how the two smoothers perform in different settings. However in practice, the motion data extracted might not always correspond to password input; for instance, it could correspond to users tapping or swiping their smartwatches to preview email, or check upcoming calendar notifications. Therefore, given the extracted motion data, we need to identify whether it is corresponding to a true password input event, or not. Snoopy addresses this by post-hoc feeding the extracted data segment into a binary classifier, which is trained on a pre-collected motion dataset covering various user interactions. In our experiments, we find that this classification step can be efficiently run on the smartwatches in real-time. Therefore, Snoopy is able to locally identify and extract the precise segments of motion data corresponding to password input events, and only send such data to the cloud for further password inference, which will be discussed in the next section. 


%% file: section/pwd_inf.tex
\chris{
\section{Deeply Learned Password Inference} 
\label{sec:pwd_inf}
As discussed in the previous section, the front-end of Snoopy runs locally on the users' smartwatches in the background, and eavesdrops the motion data (i.e. acceleration and gyroscope data) when users type or swipe their passwords. The extracted data segments corresponding to those passwords are transmitted to the cloud, where the back-end of the Snoopy system tries to infer the contents of the passwords. Snoopy has two inference models: sequence2password (seq2pwd) and sequence2digits (seq2dgt). Both models adopt a novel deep learning based password inference approach, which does not rely on accurate keystroke segmentation or handcrafted features, and is able to infer passwords reliably across different users and devices. In the following, Sec.~\ref{sub:inf_cls} first explains how we cast the problem of password inference into a classification problem. For interested readers, the background on recurrent neural networks and their use for  sequence modeling is given in the appendix. Secs.~\ref{sub:seq2pwd} and ~\ref{sub:seq2dgt} describe the design of two novel deep RNN models to infer the passwords from the captured motion data. 

\begin{figure*}[!t]
\centering
\includegraphics[width=0.7\columnwidth]{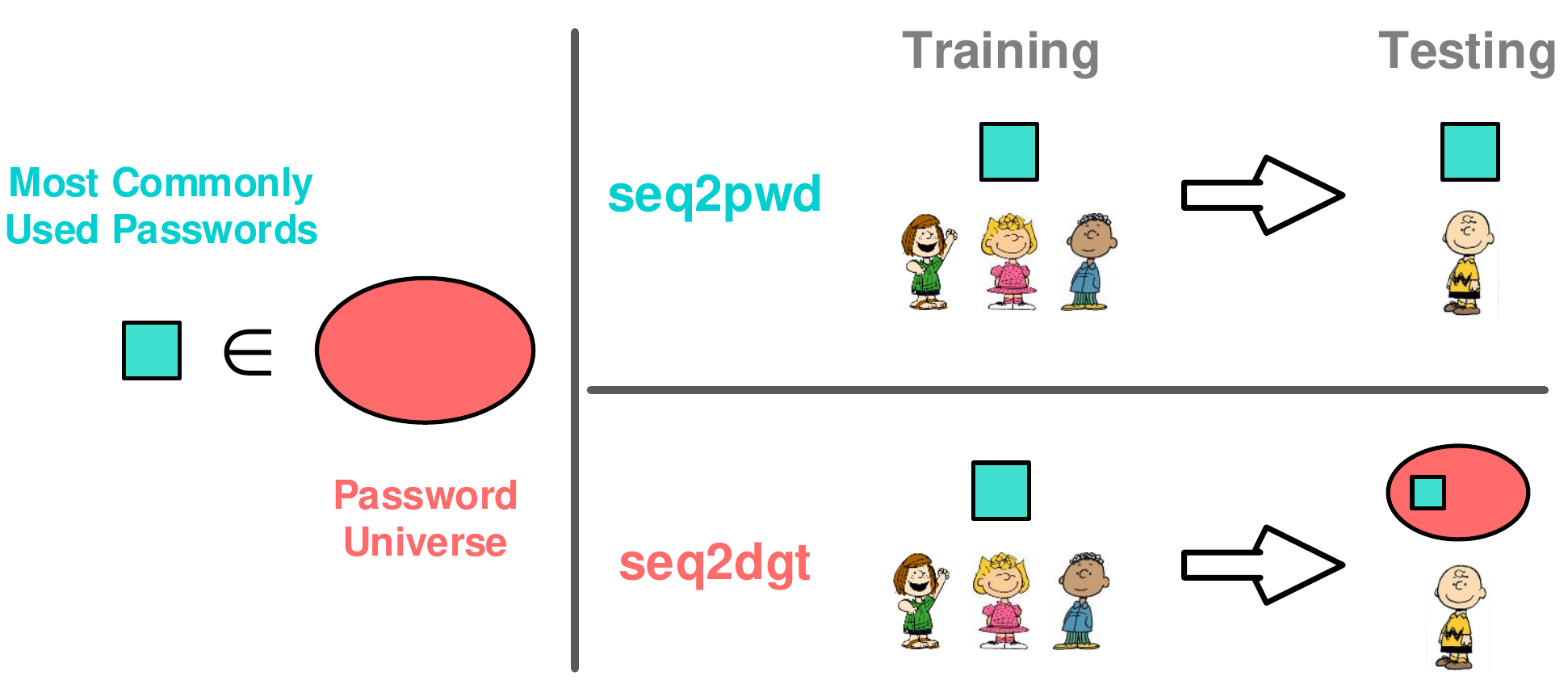}
\caption{Comparison of seq2pwd and seq2dgt models in Snoopy. Both models are able to attack users outside the training cohorts. In terms of password coverage, seq2pwd model can infer passwords seen before, while seq2dgt model is able to infer any password including those not encountered before.}
\label{fig:two_models}
\end{figure*}

\subsection{Password Inference via Classification} 
\label{sub:inf_cls}
We consider the task of password inference as a \emph{classification} problem, where the category labels are a set of passwords $P$, i.e. four digits PINs or APLs on 3$\times$3 grid. Then given the segment of motion data, the problem of inferring the password that the user has just input becomes that of finding a label within the database $P$, which can best explain the observed motion data. The size of the database $P$ determines the inference model. Though the universe of all APLs and PINs is very large, the distribution of the adoption of them in real-world is skewed. For instance, according to statistical studies~\cite{uellenbeck2013quantifying,loge2015tell}, certain APLs tend to be more popular than others, and people only use a small set of passwords due to their bias. This means that one can utilize the skewed distribution and develop their database $P$ targeting at the most commonly used passwords, which is more efficient and more cost-effective. The seq2pwd model in Snoopy is designed in this context. As in \cite{aviv2012practicality}, by taking inputs as the motion data, the proposed seq2pwd model classifies a sequence to the mostly likely passwords in $P$, without any digit segmentation. 

However, despite the high likelihood of a password existing in the most commonly used password database, the expressive power of password inference is somewhat limited as seq2pwd loses its effectiveness when encountering unseen passwords ($\notin P$). And this problem gets serious when it comes to PIN inference, as the statistics of the most commonly used PINs are not as strong as the one of APLs. To solve this problem, a seq2dgt model is also proposed in Snoopy that takes inputs as motion data but predicts the password digit-by-digit. That is to say, we could train a model by a subset of the password universe but the learnt model is able to infer any member in the universe, as long as the constituted digits are seen by the model. Fortunately, there are only 10 possibilities of the digits in APLs or PINs, which is easy to meet. Therefore, the seq2dgt is essentially a digit classifier that outputs multiple predictions at each time, where the length of predictions is decided by the password length. Notably, unlike existing work, the proposed seq2dgt model does not rely on pre-processed keystroke segmentation  \cite{miluzzo2012tapprints} and known password lengths \cite{aviv2012practicality}. It automatically learns to align the chunks of motion data to the corresponding digits and learns to predict digits without knowing in advance how many there are. This is particularly useful in the case of APLs, where keystroke segmentation is not applicable \cite{aviv2012practicality}, as the motion data of swiping APLs gives little information for digit segmentation (Fig.~\ref{fig:illus}). 

In the rest of this section, we introduce the seq2pwd model designed for APL inference as the distribution of popular adopted PINs is not as centered as that of APLs (see the survey in Sec.\ref{sec:evaluation}). The seq2dgt model is proposed for both APL and PIN inference to cover the whole universe. Fig.~\ref{fig:two_models} illustrates the inference coverage of the two models.

\begin{figure*}[t]
\centering
\begin{subfigure}[b]{0.317\textwidth}\centering
\includegraphics[width=\columnwidth]{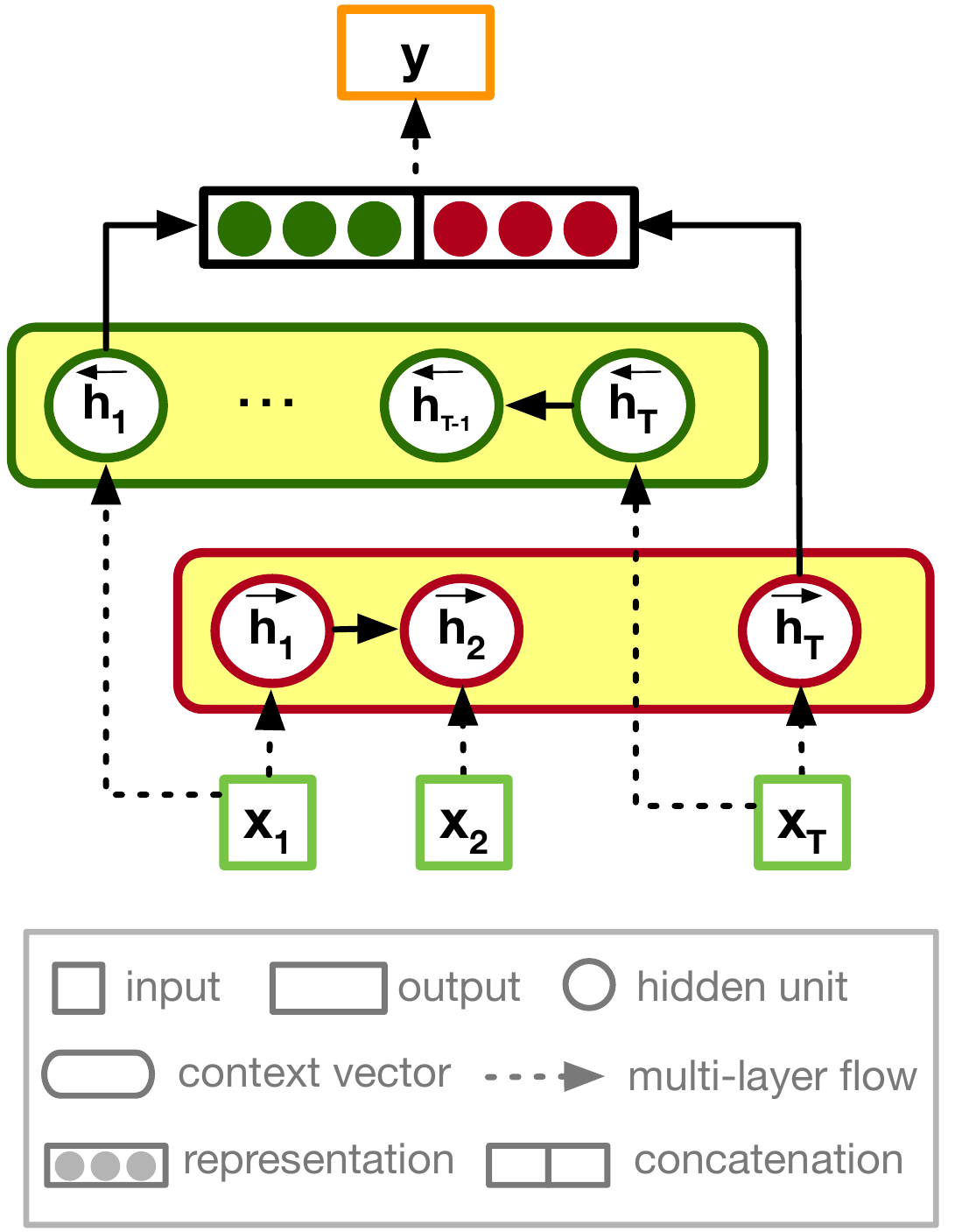} 
\end{subfigure}
\hspace*{5em}
\centering  
\begin{subfigure}[b]{0.423\textwidth}\centering
\includegraphics[width=\columnwidth]{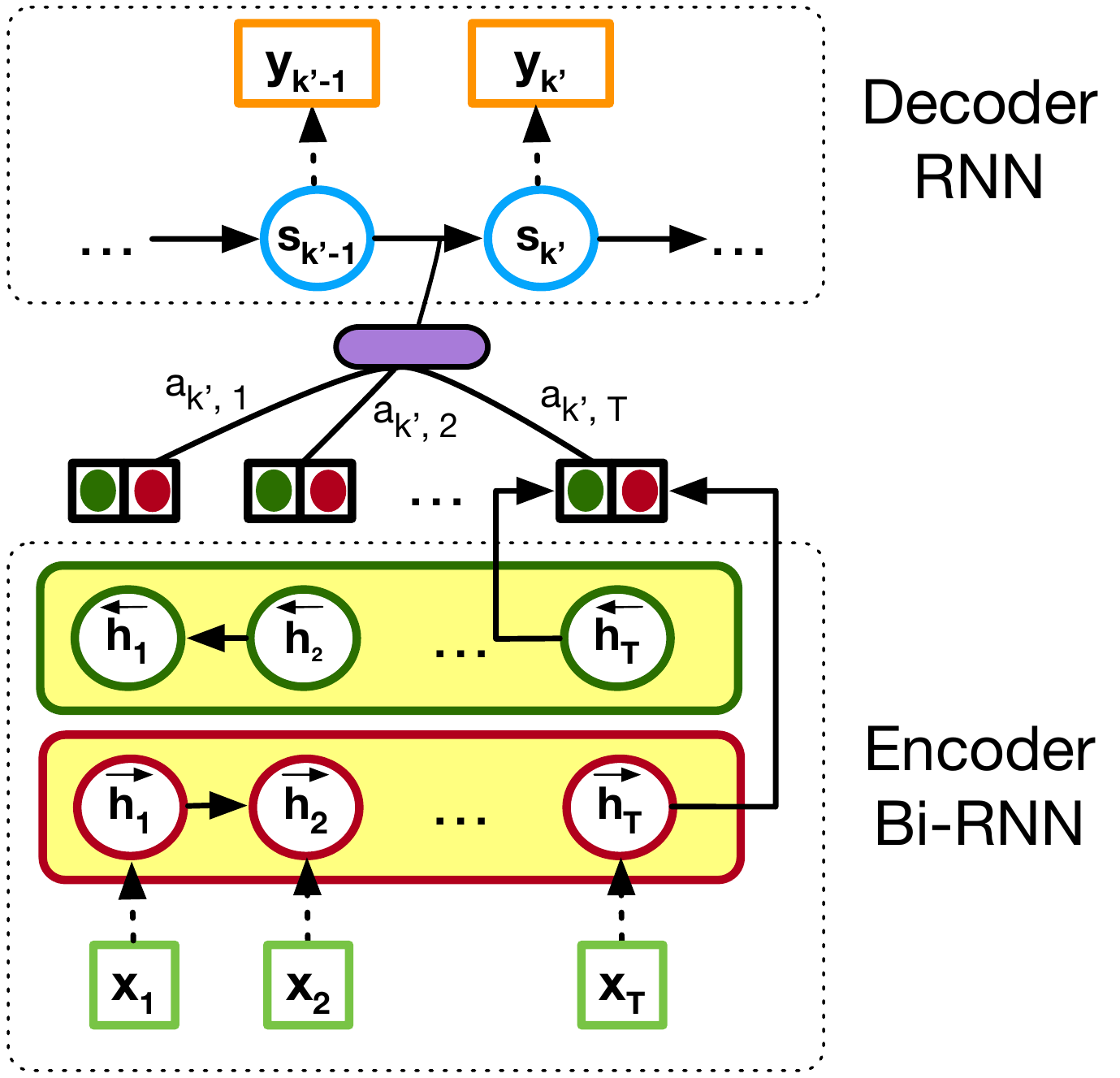} 
\end{subfigure}
\caption{The architectures of two inference models in Snoopy. Left: seq2pwd Model for commonly used password inference. Right: seq2dgt model for universal password inference.}
\label{fig:archi_rnn}
\end{figure*}

\subsection{Sequence-to-Password (seq2pwd) Model for Most Commonly Used Password Inference} 
\label{sub:seq2pwd}
Swiping passwords on the touchscreen of smartwatches, the user's finger drags the device to shift around slightly, creating slip-pulse waves in the acceleration and gyroscope data (as shown in Fig.~\ref{fig:illus}). This means the motion signals induced by swiping are more continuous than those of tapping, and typically without any gaps in between. Therefore, in this case the keystroke based inference approaches~\cite{xu2012taplogger,owusu2012accessory} won't work well since it is impossible to segment the motion data without clear boundaries between different keystrokes. 

On the other hand, for APLs the temporal correlations within the swiped pattern are much stronger, which can significantly reduce the search space and help the inference process. As discussed in Sec.\ref{sec:bkg}, given the current finger position, the smartwatch OS poses certain constraints on the possible directions of swipe \cite{uellenbeck2013quantifying}, it is only possible to swipe towards three to four reachable neighbours when starting from the top left dot. Note that although the standard RNNs with LSTMs are able to capture these correlations to a certain extent, they have certain limitations. The most important is that the network only generates output from the last hidden state (i.e. $\mathbf{h_T}$). Standard problems solved by LSTMs in NLP are with the input sequences of at most 100 samples. However, the input sequence of APLs tends to be longer, and it is more difficult for the information encoded at the beginning of the input sequence to propagate through and impact the inference results. 

To address this, Snoopy proposes a Bidirectional RNN (B-RNN) to model the rich temporal correlations within the input motion data. Concretely, at each timestamp $k$ the proposed B-RNN keeps two hidden states  $\overleftarrow{\mathbf{h}}_{k}$ and $\overrightarrow{\mathbf{h}}_{k}$, which incorporate the future ($k+1$, ..., $T$) and past (1, ..., $k-1$) information in the input sequence respectively, as shown in Fig.~\ref{fig:archi_rnn} (Left). Then B-RNN uses the same machinery to update those states from both directions. Unlike the standard network, it has two output nodes: one $\overrightarrow{\mathbf{h}}_{T}$ at the end and the other $\overleftarrow{\mathbf{h}}_{1}$ at the beginning. Therefore in B-RNN, information flows from both the start and the end of the input sequence, and the output of the network is generated from the concatenation of the two output variables $\overrightarrow{\mathbf{h}}_{T}$ and $\overleftarrow{\mathbf{h}}_{1}$. As shown in the next section, by using the bidirectional network architecture, Snoopy is able to preserve the long-term dependencies in the motion signals caused by swiping passwords, and thus infers passwords at much higher accuracy compared to competing approaches. 

\subsection{Sequence-to-Digits (seq2dgt) Model for Universal Password Inference} 
\label{sub:seq2dgt}
By formulating password inference as a sequence labeling problem, seq2pwd based RNN can guess password with a very high accuracy. However, it is difficult to adapt the seq2pwd framework to infer universal passwords. For instance, there are 389,112 possibilities for APLs and 10,000 for PINs. Using this large amount of labels for training classifiers requires huge amount of samples which is intractable in practice. We therefore propose the seq2dgt model to transform a sequence classification problem to a series of digit classification problems, where the current predicted digit conditions on the last prediction.

Despite their flexibility and power, standard RNNs can only be applied to problems whose inputs and targets share the same dimensions. It is a significant limitation in our context for two reasons. First, the lengths of APLs vary from 4 to 9, which implies the classifier needs to predict the length of digits implicitly. Fig.~\ref{fig:pwd_duration} (right) shows the duration distribution of 4 digit APLs and 7-digit APLs. As we can see, though there are 3 digits difference, the overlap of their distributions is above 40\%. Second, an input IMU readings can be as long as several hundred samples, while the readings corresponding to a certain digit are only centered in a chunk of samples. Even if the number of digits is given, associating chunks to digits is difficult, as entering PINs or APLs on smartwatches \emph{does not necessarily} occur with a uniform motion (see Fig.~\ref{fig:illus}). 

To solve the first problem, we resort to the encoder-decoder RNN architecture. It firstly uses an RNN as the encoder to map an input sequence to a context vector $\mathbf{c}$, and then stacks another RNN on it to decode the target sequence from the context vector. The decoder is often trained to predict the next sample $y_{k'}$, given the previous prediction $\{y_1, \ldots, y_{k'-1}\}$. Formally, the probability of output sequence $Y$ with the length of $T'$ (a few number of digits in our context) is defined as:
\begin{equation}
\label{equ:seq2dgt}
    p(\mathbf{Y}) = \sum_{k'=1}^{T'} p(y_{k'}|\{y_1, y_2, \cdots, y_{k'-1}\}, \mathbf{X}) 
\end{equation}
With the decoder RNN, each condition probability is modeled as:
\begin{equation}
\label{equ:decoder}
    p(y_{k'}|\{y_1, y_2, \cdots, y_{k'-1}\}, \mathbf{X})  = g(y_{k'-1}, s_{k'}, \mathbf{c})
\end{equation}
where $k$ denotes a timestep in inputs ($1<k<T$) and $k'$ is a timestep in outputs ($1<k'<T'$). $g$ is a nonlinear, potentially multi-layered function that outputs the probability of $y_{k'}$; $s_{k'}$ is the hidden state of the decoder RNN and $\mathbf{c}$ is the encoded context vector. Fig.~\ref{fig:netwrok_architecture} (Right) illustrate the above RNN architecture used in our seq2dgt model.  

By introducing a dummy digit symbol $<EOS>$, standing for end of output sequences, the unknown lengths of APLs can be implicitly determined. In this way, we have 10 candidate `digits' for each digit in APLs that our model needs to predict, i.e., $\{1,2, \ldots, 9, <EOS>\}$. The digit length of an APL is decided when the seq2dgt models gives the first $<EOS>$ symbol. For instances, a prediction $'1,2,3,6,<EOS>,9,<EOS>'$ indicates the length of target APL is $4$. All predicted digits after the first $<EOS>$ symbol e.g., $'9', <EOS>$ in the example, are not counted. $<EOS>$ usage is widely adopted in the field of NLP \cite{sutskever2014sequence}. An analogous instance of ours is machine translation, where a source English sentence may not have the same number of words as its Chinese translation. The dummy $<EOS>$ symbol can prevent the model generating an infinite number of words. Note that, this step is only for APLs; a PIN's length is fixed to 4 in most scenarios. 

Originally, the context vector $\mathbf{c}$ is computed by encoding all inputs. However, the second problem remains as the IMU readings are sampled at 200Hz, whose lengths are dramatically longer than a few digits but only a part of them contribute at one decoding timestep. Here we introduce the attention mechanism in our seq2dgt model. Formally, the conditional probability in this attention seq2dgt is defined as:
\begin{equation}
\label{equ:attention_decoder}
    p(y_{k'}|\{y_1, y_2, \cdots, y_{k'-1}\}, \mathbf{X})  = g(y_{k'-1}, s_{k'}, \mathbf{c}_{k'})
\end{equation}
Unlike the conditional probability in Eq.~(\ref{equ:decoder}), here the probability is conditioned on a distinct context vector $\mathbf{c}_{k'}$ for each output digit $y_{k'}$. The new context vector depends on a sequence of hidden sates ($h_1, \ldots, h_T$) to which an encoder maps the input sequence $\mathbf{X}$, where we adopt a bidirectional RNN, i.e., $h_k = [\overleftarrow{\mathbf{h}}_{k}, \overrightarrow{\mathbf{h}}_{k}]$. Formally,
\begin{equation}
\label{equ:ctx_vector}
    \mathbf{c}_{k'} = \sum_{k=1}^{T} a_{k'k}h_k
\end{equation}
where $a_{k'k}$ are the weights determing the contribution of $h_k$ in encoding $\mathbf{c}_{k'}$ for the $k'$-th digit and it can be determined through backpropagation in an end-to-end optimization. The attention mechanism is widely adopted in the scenario where input sequences are very long, and a single context vector is too compressed to decode outputs. For example, Hermann et al. \cite{hermann2015teaching} have achieved impressive results in document summarization by introducing the attention mechanism in their models, which solves the problem that the number of words of in the input documents are much larger than the ones in the output summaries. As shown in the next section, the seq2dgt model benefits from this attention mechanism and it is able to adaptively focus on specific chunks of input (with high attention weights) when generating digits in different positions of the passwords. 

\begin{figure*}[t]
\centering
\begin{subfigure}[b]{0.33\textwidth}\centering
\includegraphics[width=\columnwidth]{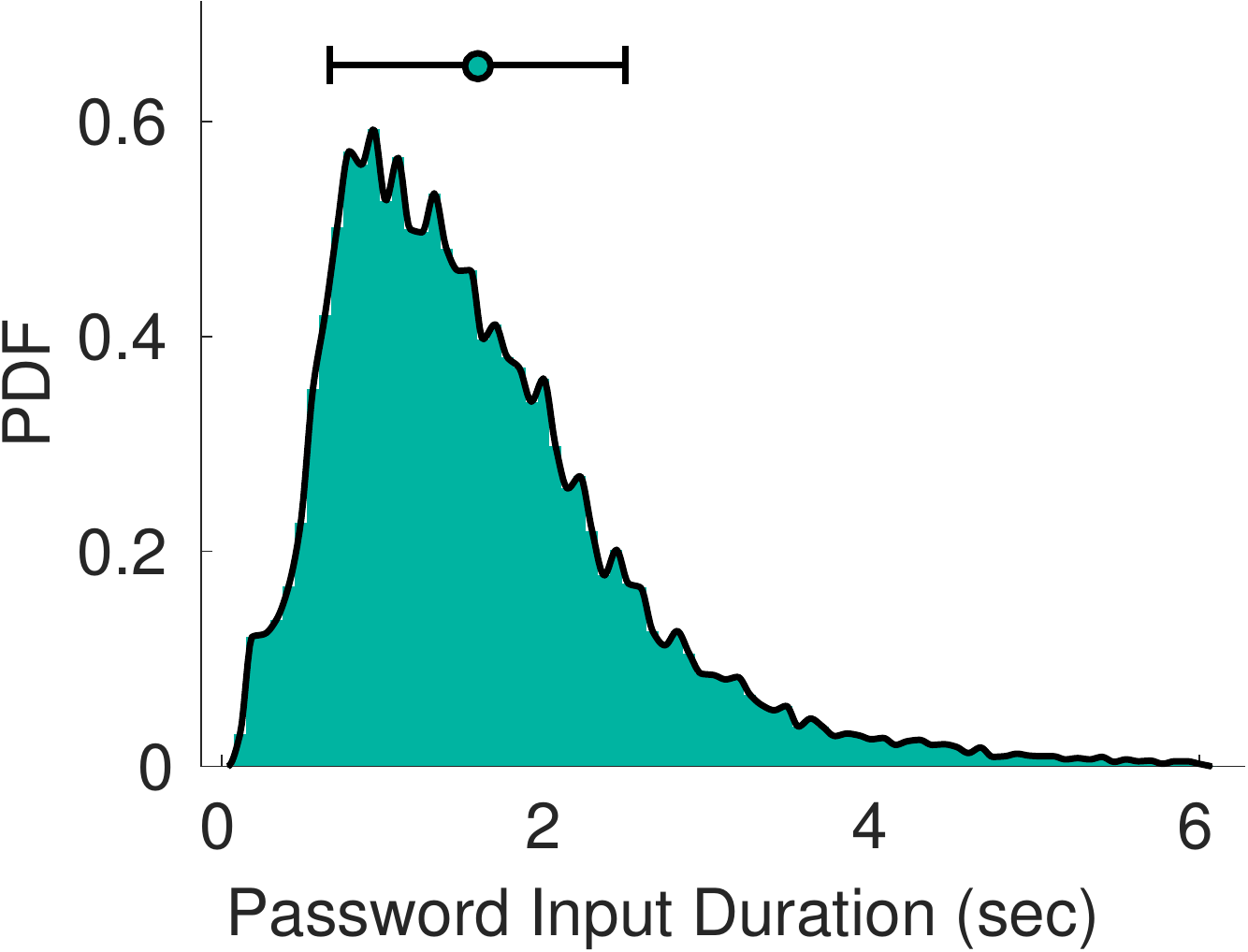}  
\end{subfigure}%
\hspace*{4em}
\centering
\begin{subfigure}[b]{0.33\textwidth}\centering
\includegraphics[width=\columnwidth]{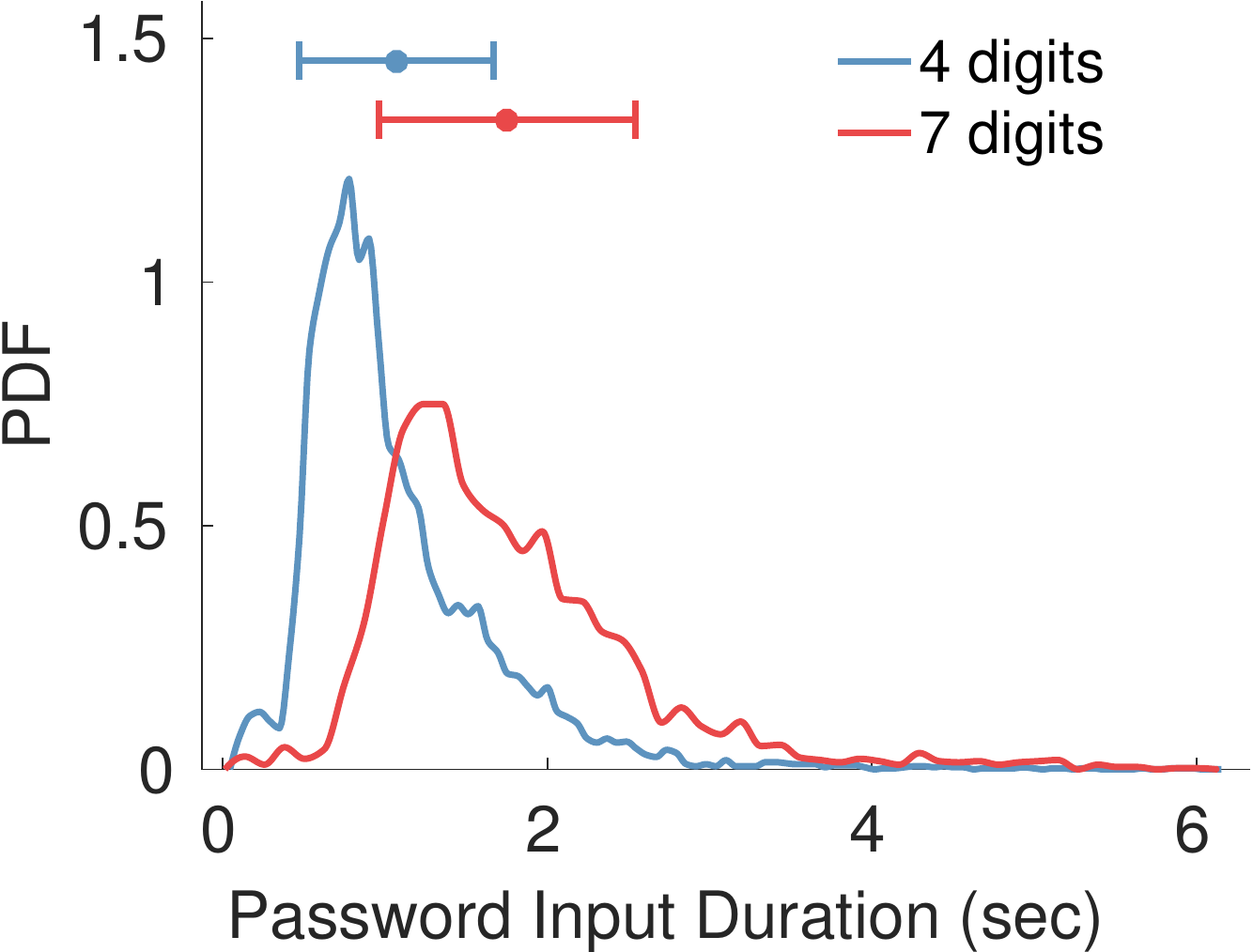} 
\end{subfigure}%
\caption{PDF of APL input duration. Left: duration distribution of swiping APLs. Right: Differentiating a 4-digit APL and 7-digit APL is difficult based on their duration distribution.}
\label{fig:pwd_duration}
\end{figure*}

}

%% file: section/eval.tex
\section{Evaluation} 
\label{sec:evaluation}
\hongkai{We evaluate the proposed Snoopy system extensively on large-scale real world datasets collected in three different sites: \emph{Oxford}, \emph{Shanghai} and \emph{Harbin}. In total, our experiments collected preferred/generated passwords from \textbf{420} anonymous participants, recruited a separate group of \textbf{362} volunteers to contribute their motion data when entering passwords on smartwatches (worn on their left hands), and accumulated over \textbf{60$k$} samples of motion data during password entries\footnote{The study has received ethical approval \emph{R50768} from the University of Oxford.}. Our code and dataset are available in \url{https://github.com/ChristopherLu/snoopy/}. 

In the following, Sec.~\ref{sub:eval-detection} focuses on evaluating the performance of the front-end password extraction capability of the proposed Snoopy system; Sec.~\ref{sub:apl-eval} presents the performance of Snoopy in inferring APLs; and finally Sec.~\ref{pin-eval} assesses our system's capability of PIN inference. 
}

\input{section/eval_detection}
\input{section/eval_inference}

%% file: section/eval_detection.tex
\subsection{Performance of Password Input Extraction} 
\label{sub:eval-detection}

\subsubsection{Experimental Setup}
\label{ssub:d-setup}
\hfill \break
\noindent \textbf{Data Collection:}
To evaluate the performance of password input extraction, we recruited $15$ volunteers ($10$ males and $5$ females), and asked them to wear different smartwatches (Android and iWatches) on their left wrists. Throughout our experiments we use four different models of Android watches: Sony SmartWatch3, Samsung Gear Live, Moto 360 Sports, LG Urbane, and two Apple watches: 38mm and 42mm versions of iWatch2. Note that Android smart watches typically use APLs for system level authentication, whereas Apple watches use PINs. 

During the experiments, we asked the participants to perform three different types of actions: \emph{password input} where they enter their passwords on their smartwatches; \emph{non-password input} where they tap/swipe on the watches screen to do other tasks (e.g. preview email or check a calendar notification) but not to enter a password; and \emph{no input} when they just perform a series of activities wearing their smartwatches, such as drinking, drawing, eating, walking, going down/upstairs, typing on keyboards and holding hands still. We designed a data collection app on the smartwatches, which samples the motion sensors in the background (100Hz in this case), and instructs the participants to perform certain actions at a given time. In this way we can obtain accurate ground truth as to when the user is performing a certain action.

To collect rich enough data, in one episode we requested a participant to at least perform three actions, where the action in the middle should be password input or non-password input action, e.g. she may first walk, then enter her password, and finally go upstairs. Each participant is requested to contribute multiple episodes, and in total we obtained 455 episodes for Android watches, and 387 for Apple watches. 

\noindent \textbf{Competing Approaches: }
Since the front-end of Snoopy has to run locally, in this series of experiments we only consider lightweight approaches that can run in real-time on the smartwatches. Recall that the task of password extraction has not been attempted on smartwatches before. Previous related work has focused on the extraction of PINs on smartphones only~\cite{xu2012taplogger,cai2012practicality,miluzzo2012tapprints}. However, features used to extract keystrokes on smartphones are not suitable for smartwatches due to the limited screen size and low signal to noise ratio (see Fig.~\ref{fig:snr}). And even in the case of smartphones, previous work has assumed that any keystroke is part of a password; however in practice, keystrokes could be used in the middle of other tasks not related to password input, for example replying to email. There is no competing approach that currently addresses the entire password extraction task. In what follows, we evaluate a realistic password extraction approach for smartwatches that has three stages. In the first stage, we assess how well we can detect the beginning of a password (see Sec.~\ref{sub:pwd_detection} for details); we compare a number of classifiers including the Support Vector Classifier (\textbf{SVC}), \textbf{decision trees},  \textbf{logistic regression}, \textbf{naive Bayesian} and \textbf{random forest} classifiers. Once we have detected the beginning of a password, in the second stage, we continue to define the full extent of the potential password, by classifying individual frames, and smoothing classification results (see Sec.~\ref{sub:seq_smoothing} for details). Here for smoothing, we compare the Hidden Markov Model (\textbf{HMM}) based and the voting based moving average (\textbf{moving average}) approaches implemented in Snoopy with the baseline approach without smoothing (\textbf{raw}). Once the start and end of a potential password are found, the final stage classifies this sequence as an actual password, or a non-password sequence (as discussed at the end of Sec.~\ref{sub:seq_smoothing}); here we also compare the performance of several classifiers, including \textbf{SVC}, \textbf{decision trees}, \textbf{logistic regression}, \textbf{naive Bayesian} and \textbf{random forest}. The collected dataset is split into a training set (data from 10 subjects) and a test set (data from the other 5 subjects), and we consider 5-fold cross-validation.

\subsubsection{Experiment Results}
\label{ssub:d-results}
\begin{figure*}[t]
\centering
\begin{subfigure}[b]{0.45\textwidth}\centering
\includegraphics[width=\columnwidth]{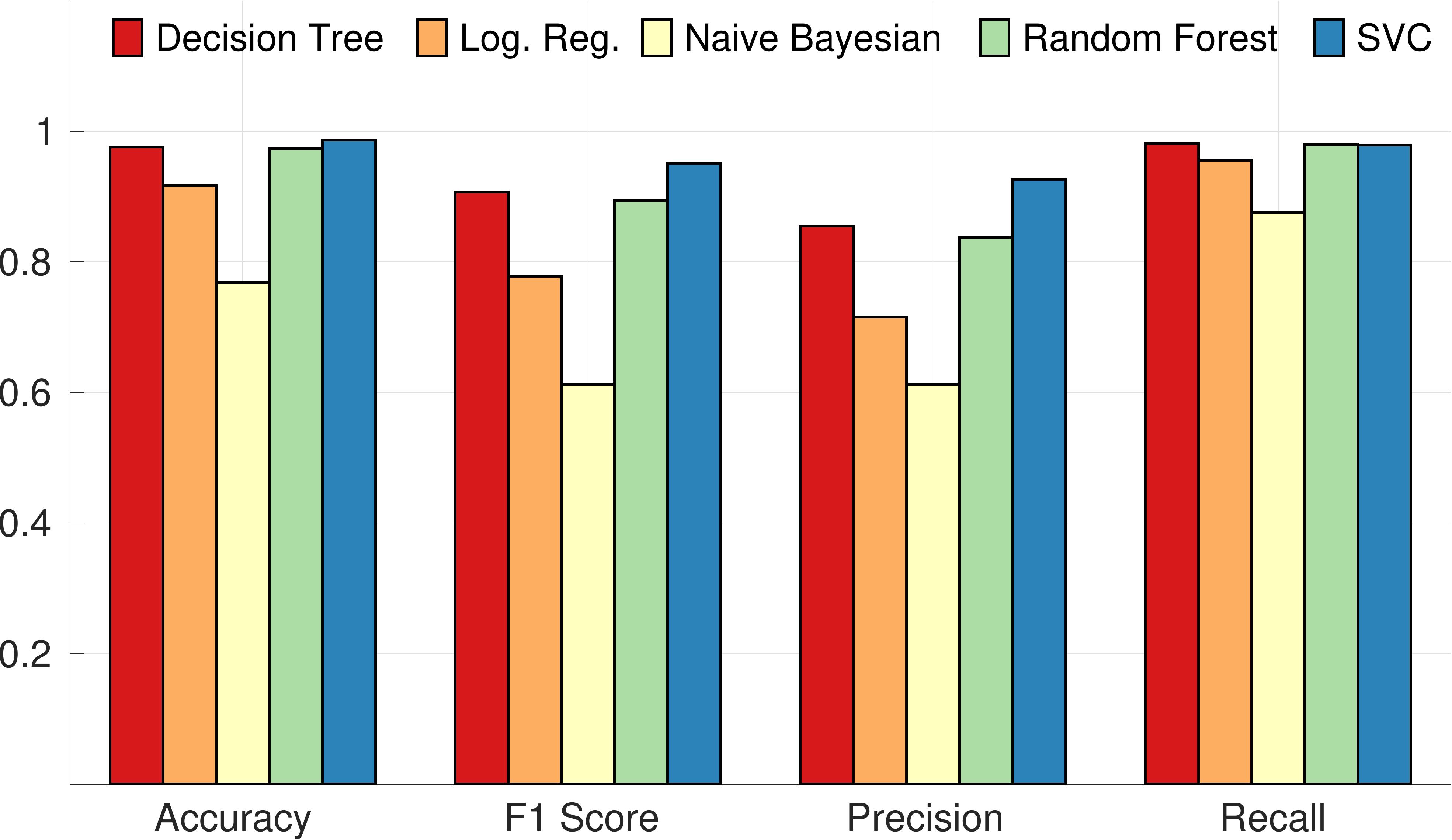}  
\end{subfigure}%
\hfill
\centering
\begin{subfigure}[b]{0.45\textwidth}\centering
\includegraphics[width=\columnwidth]{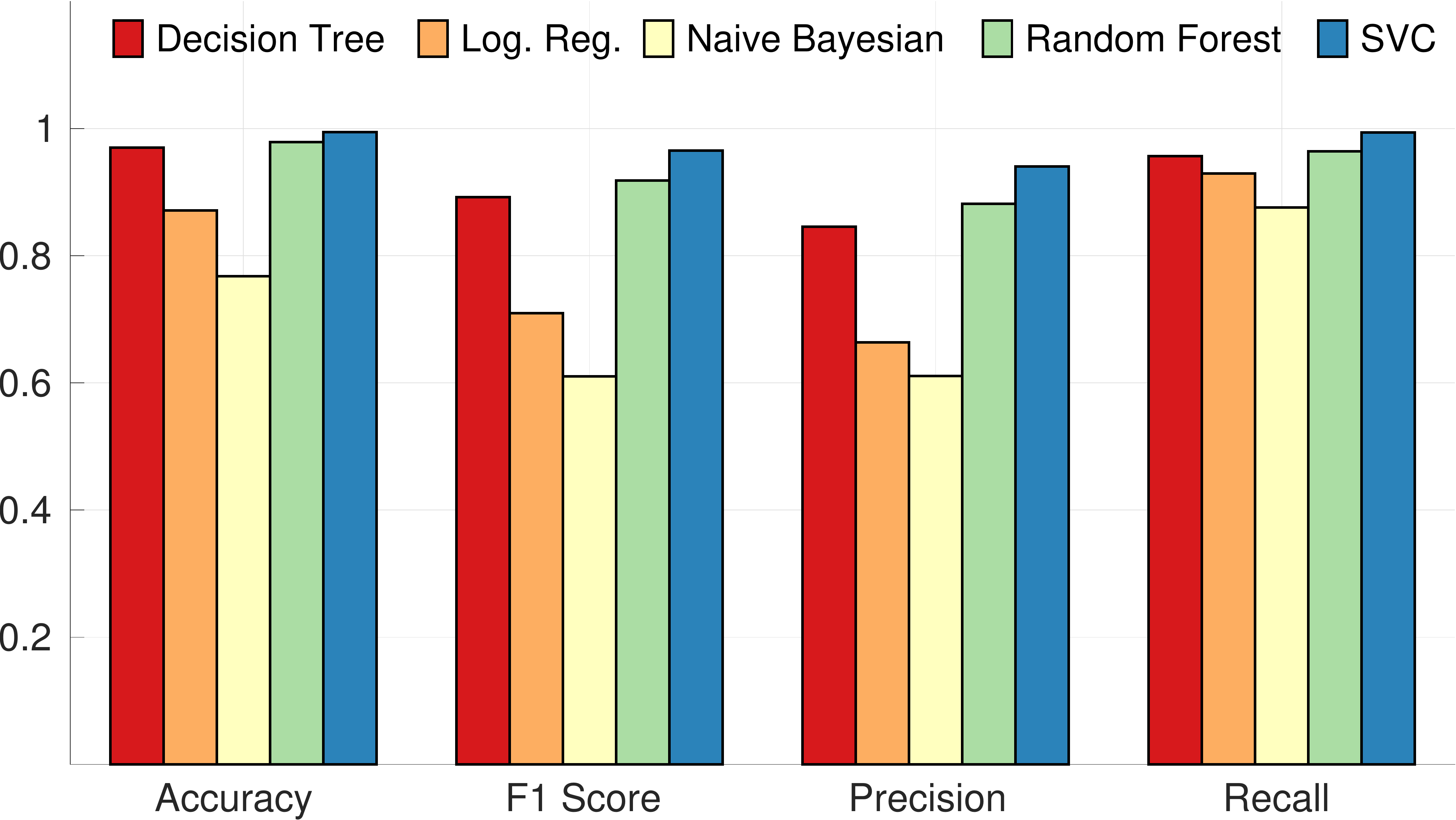} 
\end{subfigure}%
\caption{Performance of detecting potential password input events. Left: PINs; Right: APLs}
\label{fig:bar_detection}
\end{figure*}

\hfill \break
\noindent \textbf{Detecting Password Input Events: }
As discussed in Sec.~\ref{sub:pwd_detection}, for a given frame of acceleration data, we would like to decide whether it corresponds to password input or not. As in many other binary classification problems, here we consider the precision, recall, $F_1$ score and accuracy of the classifiers. Fig.~\ref{fig:bar_detection} shows that the SVC outperforms competing classifiers in terms of all evaluation metrics. For both PINs and APLs, it can achieve $>0.95$ $F_1$ scores and $>0.98$ accuracy. The random forest and decision tree classifiers can achieve comparable recall with SVC, but their precision is nearly $8\%$ lower than that of SVC. Based on these results, in what follows, we adopt SVC as the default classifier for detecting the beginning of a potential password input event in Snoopy. 

\begin{figure*}[t]
\centering
\begin{subfigure}[b]{0.45\textwidth}\centering
\includegraphics[width=\columnwidth]{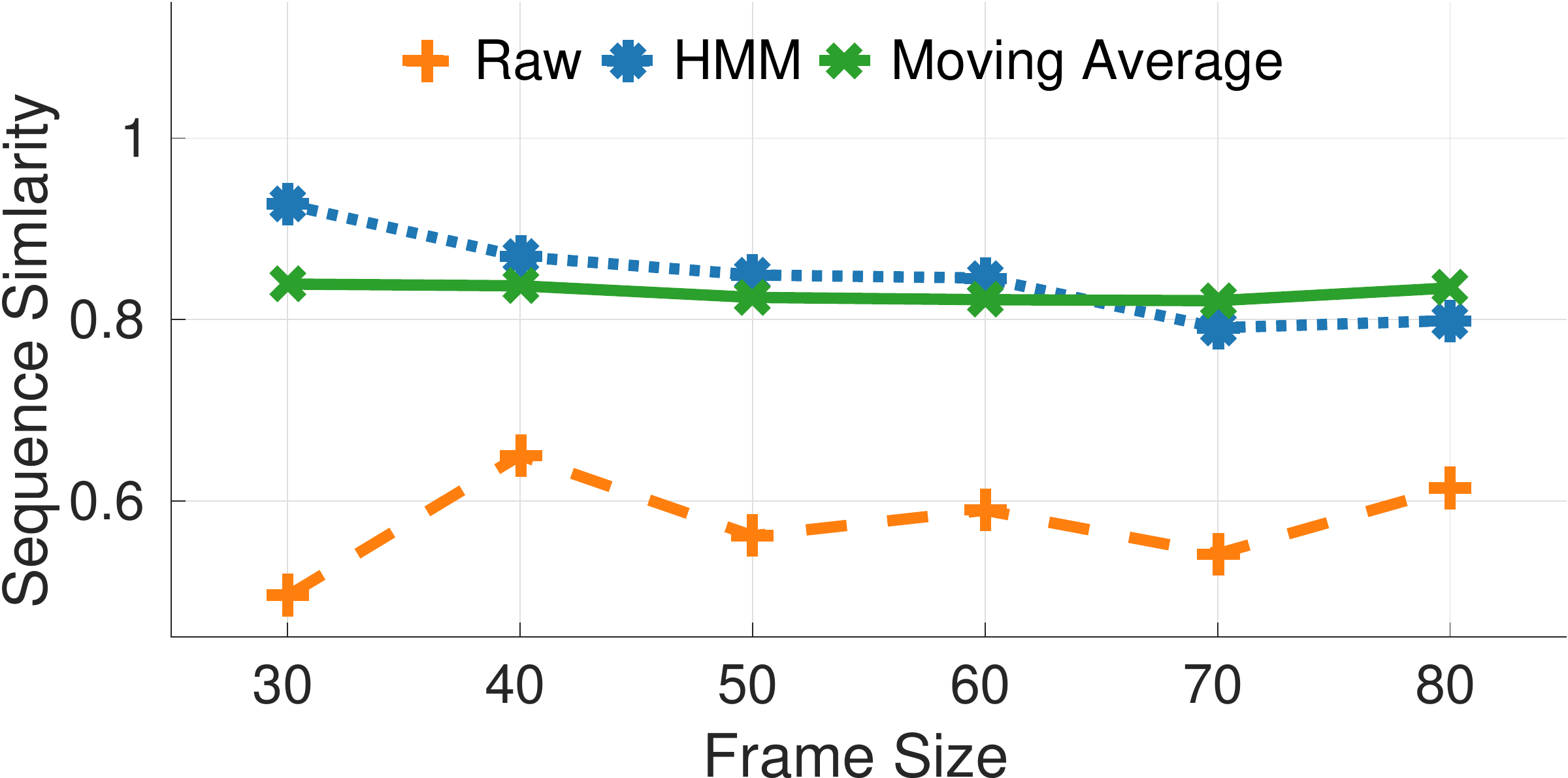}  
\end{subfigure}%
\hfill
\begin{subfigure}[b]{0.45\textwidth}\centering
\includegraphics[width=\columnwidth]{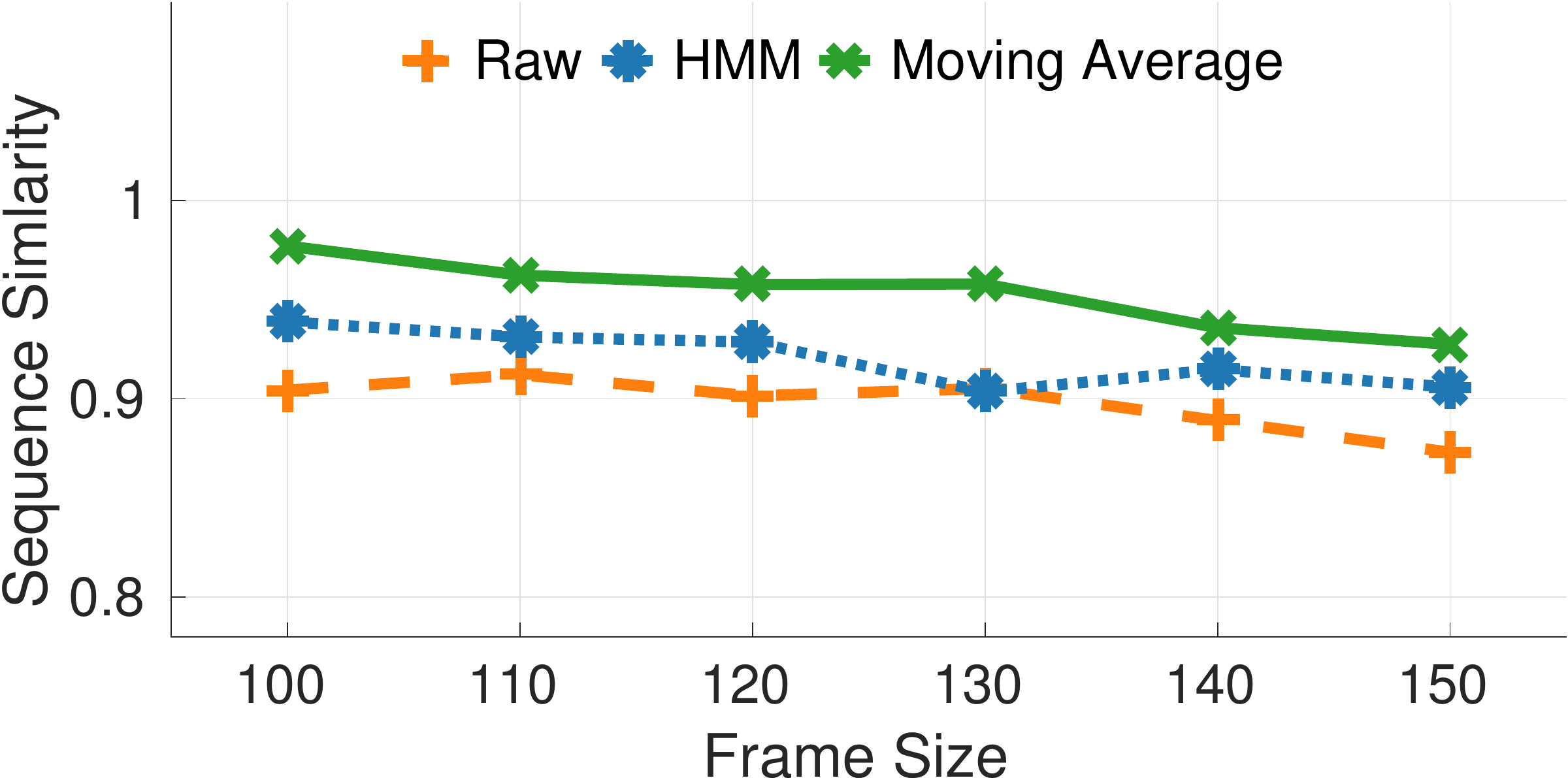} 
\end{subfigure}%
\caption{Performance of sequence smoothing when using different frame sizes. Left: PINs; Right: APLs}
\label{fig:line_segmentation}
\end{figure*}

\noindent \textbf{Smoothing Detected Sequences: }
The above SVC approach is iteratively used in ensuing frames to classify them as password-related (`1') or not (`0'). When we start seeing a lot of `0'-s this indicates the end of a potential password. The candidate password sequence that we derive (e.g. `11001111110001101') is then passed through a smoothing process as discussed in Sec.\ref{sub:seq_smoothing}. The smoother adjusts the frame labels taking into account those of nearby frames, and further refines the start and end point of the potential password input event. To evaluate the performance of the smoothing process, we consider the similarity of a frame sequence $s_r$ with respect to the ground truth $s_t$: 
\begin{equation}
\label{eqn:seq_similarity}
  d(s_r, s_t) = \frac{|s_r \cap s_t|}{(|s_r|+|s_t|)/2}
\end{equation} 
where $|\cdot|$ is the cardinality of the positive labels, and $|s_r \cap s_t|$ is the number of frames that have the same labels in both $s_r$ and $s_t$. Intuitively a sequence $s_r$ with higher $d$ is better because it is closer to the ground truth, and thus tends to contain larger portion of correctly labelled frames. Fig.~\ref{fig:line_segmentation} shows the average similarity scores of sequences generated by different smoothers (HMM and majority voting moving average vs. no smoothing). As we can see for different frame sizes, sequences without smoothing (raw) consistently have lower $d$ scores. This confirms that the smoothing process is beneficial. For instance, for APLs, the moving average smoother can reach 0.98 in terms of sequence similarity score, while for PINs the HMM smoother can achieve 0.94. Note that for the PINs, the performance gain between not using (raw) and using smoothers (HMM and moving average) can be up to 35\%. This is because the motion data generated by PINs contains ``gaps'' between two adjacent finger touches, where the detection approaches would naturally label frames within the gaps as negative (i.e. no password input). In those cases, the smoothers can correct those errors, and output a sequence with more consistent labels. It is also interesting to see that although the two smoothers considered in Snoopy have comparable performance, HMM tends to work better than moving average for PINs, but can be inferior for APLs. Again this is because HMM is able to mitigate those non-informative gaps within data of the PINs, while for APLs moving average is more robust.

\noindent \textbf{Password-positive vs. Password-negative Sequences: }
Given a smoothed sequence, the front-end of Snoopy needs to identify if it corresponds to an actual password input, or non-password input such as swiping to check notifications. As discussed in Sec.\ref{sub:seq_smoothing}, Snoopy addresses this by feeding an entire sequence into a binary classifier. Tables~\ref{tab:pin_post} and~\ref{tab:apl_post} show the identification performance of different classifiers for PINs and APLs respectively. We see that unlike the more expensive random forest and SVC, the simpler classifiers work surprisingly well for this task. Note that for both types of passwords, the simple classifiers can achieve up to $>$0.95 $F_1$ score, which means they are able to reliably distinguish motion caused by password input from that of other interactions. This is because due to the small size of the smartwatches, the ways to interact with their touchscreens are very limited. Therefore the motion signals of PINs or APLs are very unique compared to others. 
\begin{table}
\centering
\makebox[0pt][c]{\parbox{1\textwidth}{%
    \begin{minipage}[b]{0.36\hsize}\centering
        \small
        \begin{tabular}{|c|c|c|c|}
\hline
                                                              & \begin{tabular}[c]{@{}c@{}}F1\\ Score\end{tabular} & Precision & Recall \\ \hline
\begin{tabular}[c]{@{}c@{}}Decision\\ Tree\end{tabular}       & 0.89                                               & 0.92      & 0.86   \\ \hline
\begin{tabular}[c]{@{}c@{}}Naive\\ Bayesian\end{tabular}      & 0.91                                               & 0.90      & 0.91   \\ \hline
\begin{tabular}[c]{@{}c@{}}Logistic\\ Regression\end{tabular} & 0.95                                               & 0.96      & 0.93   \\ \hline
SVC                                                           & 0.93                                               & 0.92      & 0.95   \\ \hline
\end{tabular}
        \caption{\textbf{PIN} sequence identification results.}
        \label{tab:pin_post}
    \end{minipage}    
    \begin{minipage}[b]{0.36\hsize}\centering
        \small
        \begin{tabular}{|c|c|c|c|}
\hline
                                                              & \begin{tabular}[c]{@{}c@{}}F1 \\ Score\end{tabular} & Precision & Recall \\ \hline
\begin{tabular}[c]{@{}c@{}}Decision\\ Tree\end{tabular}       & 0.93                                                & 0.91      & 0.93   \\ \hline
\begin{tabular}[c]{@{}c@{}}Naive\\ Bayesian\end{tabular}      & 0.91                                                & 0.90      & 0.91   \\ \hline
\begin{tabular}[c]{@{}c@{}}Logistic\\ Regression\end{tabular} & 0.96                                                & 0.97      & 0.95   \\ \hline
SVC                                                           & 0.94                                                & 0.95      & 0.94   \\ \hline
\end{tabular}
        \caption{\textbf{APL} sequence identification results.}
        \label{tab:apl_post}
    \end{minipage}
    \hfill
    \begin{minipage}[b]{0.26\hsize}\centering
        \small
        \begin{tabular}{|c|c|c|}
\hline
\textbf{\begin{tabular}[c]{@{}c@{}}ave./max\\ (\%)\end{tabular}} & \begin{tabular}[c]{@{}c@{}}Feature\\ Extraction\end{tabular} & SVM       \\ \hline
\begin{tabular}[c]{@{}c@{}}Sony\\ SW3\end{tabular}               & 8.9/23.0                                                     & 7.2/21.4  \\ \hline
\begin{tabular}[c]{@{}c@{}}Samsung\\ Gear Live\end{tabular}      & 10.9/25.2                                                    & 9.8/18.5  \\ \hline
\begin{tabular}[c]{@{}c@{}}Moto 360\\ Sports\end{tabular}        & 10.5/22.7                                                    & 10.1/25.5 \\ \hline
\end{tabular}
        \caption{CPU load of running feature extraction and SVM.}
        \label{tab:svc_costs}
    \end{minipage}    
}}
\end{table}

\noindent \textbf{Resource Consumption: }
\hongkai{The final set of experiments analyse the resource consumption of Snoopy front-end on three Android watches with different hardware specifications. Tab.~\ref{tab:sys_perf} shows the average/maximum CPU load, delta current consumption and battery usages when the front-end is running the following three tasks: a) password input detection (Sec.~\ref{sub:pwd_detection}); b) password-input data smoothing and identification (Sec.~\ref{sub:seq_smoothing}); and c) uploading extracted data to the back-end. We see that among the three tasks, uploading actually consumes the largest amount of energy, while detection and smoothing are relatively cheap. On the other hand, detection and smoothing tend to occupy the CPU more than uploading. This is expected because it is well known that transmitting over WiFi is power-consuming on smartwatches, while detection and smoothing are more computation-intensive as they involve running SVM classifiers. More specifically, as shown in Tab.~\ref{tab:svc_costs}, on all three watch platforms, running feature extraction and SVM classifiers consume similar level of CPU resources ($\sim$10\%), but the former is slightly more expensive since it involves continuous caching operations, e.g. maintaining the sliding windows. In addition, like many other apps~\cite{liu2016lasagna}, to minimize impact on battery lifetime, Snoopy only uploads cached data when the watches are connected to power with WiFi connections available. As shown in Tab.~\ref{tab:sys_perf}, in general the Snoopy front-end doesn't require excessive resources, and when disguised as an innocent fitness app, it is not likely to have noticeably abnormal energy/computation impact on the smartwatches. }

\begin{table}[!ht]
\small
\centering
\begin{tabular}{|c|c|c|c|c|c|c|c|}
\hline
\textbf{Model}                                                               & \textbf{SoC}                                                                          & \textbf{RAM}           & \textbf{\begin{tabular}[c]{@{}c@{}}Battery \\ Cap.\end{tabular}} & \textbf{Task} & \textbf{CPU load (avg/max)} & \textbf{Current delta} & \textbf{\begin{tabular}[c]{@{}c@{}}Battery\\ Usage\end{tabular}} \\ \hline
\multirow{3}{*}{Sony SW3}                                                    & \multirow{3}{*}{\begin{tabular}[c]{@{}c@{}}Qualcomm\\ APQ8026\\ SD 400\end{tabular}}  & \multirow{3}{*}{512MB} & \multirow{3}{*}{420 mAh}                                         & Detection     & 9.2\%/17.5\%                 & 8.9 mA                 & \multirow{3}{*}{2\% per hr}                                      \\ \cline{5-7}
                                                                             &                                                                                       &                        &                                                                  & Smoothing   & 7.2\%/15.2\%                & 3.1 mA                 &                                                                  \\ \cline{5-7}
                                                                             &                                                                                       &                        &                                                                  & Uploading          & 2.4\%/9.9\%                 & 18.9 mA                &                                                                  \\ \hline
\multirow{3}{*}{\begin{tabular}[c]{@{}c@{}}Samsung\\ Gear Live\end{tabular}} & \multirow{3}{*}{\begin{tabular}[c]{@{}c@{}}Qualcomm\\ MSM8226\\ SD 400\end{tabular}}  & \multirow{3}{*}{512MB} & \multirow{3}{*}{300 mAh}                                         & Detection     & 8.1\%/19.4\%                & 11.4 mA                & \multirow{3}{*}{3\% per hr}                                      \\ \cline{5-7}
                                                                             &                                                                                       &                        &                                                                  & Smoothing   & 15.6\%/29.3\%               & 6.2 mA                 &                                                                  \\ \cline{5-7}
                                                                             &                                                                                       &                        &                                                                  & Uploading          & 2.1\%/19.3\%                & 25.1 mA                &                                                                  \\ \hline
\multirow{3}{*}{\begin{tabular}[c]{@{}c@{}}Moto 360\\ Sports\end{tabular}}   & \multirow{3}{*}{\begin{tabular}[c]{@{}c@{}}Qualcomm \\ MSM8926\\ SD 400\end{tabular}} & \multirow{3}{*}{512MB} & \multirow{3}{*}{300 mAh}                                         & Detection     & 8.3\%/26.2\%                & 16.1 mA                & \multirow{3}{*}{3\% per hr}                                      \\ \cline{5-7}
                                                                             &                                                                                       &                        &                                                                  & Smoothing   & 7.3\%/22.4\%                & 3.8 mA                 &                                                                  \\ \cline{5-7}
                                                                             &                                                                                       &                        &                                                                  & Uploading          & 2.3\%/26.1\%                & 22.3 mA                &                                                                  \\ \hline
\end{tabular}
\caption{Resource consumption (CPU and power) of the Snoopy front-end on smartwatches with different hardware specs.}
\label{tab:sys_perf}
\end{table}

%% file: section/eval_inference.tex
\subsection{Performance of APL Inference} 
\label{sub:apl-eval}
We are now in a position to turn our attention to how the back-end password inference component of Snoopy performs. In this section we firstly discuss the performance of APL inference, while the PIN inference will be covered in Sec.~\ref{pin-eval}.

\subsubsection{Experiment Setup} 
\label{ssub:apl-setup}
\hfill \break
\hongkai{\noindent \textbf{APL Database Construction: }
As discussed in Sec.~\ref{sec:pwd_inf}, to infer the user entered APLs, both the seq2pwd and seq2dgt models considered in Snoopy require a good password database $P$ for training, which can cover as many common passwords as possible. To construct such a database $P$, we consider the publicly available APL data reported in~\cite{loge2015tell} and also collected our own dataset. The APL dataset in ~\cite{loge2015tell} contains $\sim$$4,000$ APLs entries collected from the anonymous users (with duplications). From this dataset, we rank the distinct APLs according to their frequencies, and select the most popular 113 APLs that can cover half of all the APL entries (2000 out of 4000). This ensures that the selected APLs can achieve a good coverage of the most commonly used APLs, while leaving out those APLs that are seldom used. 

We also recruited $112$ anonymous participants to survey their preferred passwords (both PIN and APL) when using mobile devices. The purpose of collecting our own password dataset is to obtain an independent dataset in addition to the publicly available data, which would make the constructed password database $P$ more diverse. During the data collection process, we have made sure that every step complied with data privacy policies, and there is no link between the collected data and any individual participant. In particular, we have first obtained the participants' consent that their data will be used in a scientific study to evaluate password security on smartwatches. If a participant agreed to proceed, she was then given an Android watch, and we asked her to wear the watch on her left wrist. Then the participant is provided with an instruction sheet, which asks her to set a password in a survey app on the smartwatch. The survey app only records the entered passwords by monitoring touches on the touchscreen. When the participant finished entering the password, it asks if she is aware of the purpose of the study and would like to contribute this password. If so, the password is assigned with a unique random ID, and written into a random line of a local text file on the smartwatch. Otherwise there is no information saved. Note that during this process, the participants were asked to input passwords in private and take their time. The watches and instruction sheet were passed directly to the next participant without our intervention. After the survey process, fortunately we obtained $112$ APL entries from all participants, among which we have extracted $64$ distinct APLs. Finally, we fuse those $64$ surveyed APLs and the $113$ APLs extract from the existing dataset~\cite{loge2015tell}, and construct a password database $P$ with $147$ \emph{distinct} APLs. 

\noindent \textbf{APL Input Motion Data Collection: }
Given the above constructed APL database $P$, we recruited a total number of $322$ participants across three experiment sites to collect the motion data when they are entering APLs on their smartwatches. Each participant was randomly given 6 APLs selected from $P$. The participants were asked to wear the smartwaches on their left wrist but in the most comfortable way, and then enter each password in our data collection app about 20 times. The app logs the ground truth by monitoring tap/swipe on the smartwatch screen, and saves the motion data at the same time. In total, we have collected $36,569$ valid samples, each of which contains an APL and the motion data when it was entered. This set of data is used to train our models in Snoopy. 
}

\noindent \textbf{Competing Approaches: }
\hongkai{We implement both the seq2pwd and seq2dgt models considered in Snoopy using Keras~\cite{chollet2015keras}, and train them on NVIDIA K80 GPUs with the Adam optimiser~\cite{kingma2014adam}. To the best of our knowledge, Snoopy is the first work to study the problem of inferring smartwatch APLs, and there is no existing work that can infer APLs without knowing the exact segmentation of digits within APLs (as discussed in~\ref{sub:seq2dgt}). Therefore, here we only consider one of the best APL inference approach designed for smartphones, \textbf{GestureLogger}~\cite{aviv2012practicality}, which bears some resemblance to the proposed seq2pwd model in Snoopy.}

\begin{figure*}[t!]
  \begin{minipage}[t]{0.47\textwidth} 
  \centering   
  \includegraphics[width=\columnwidth]{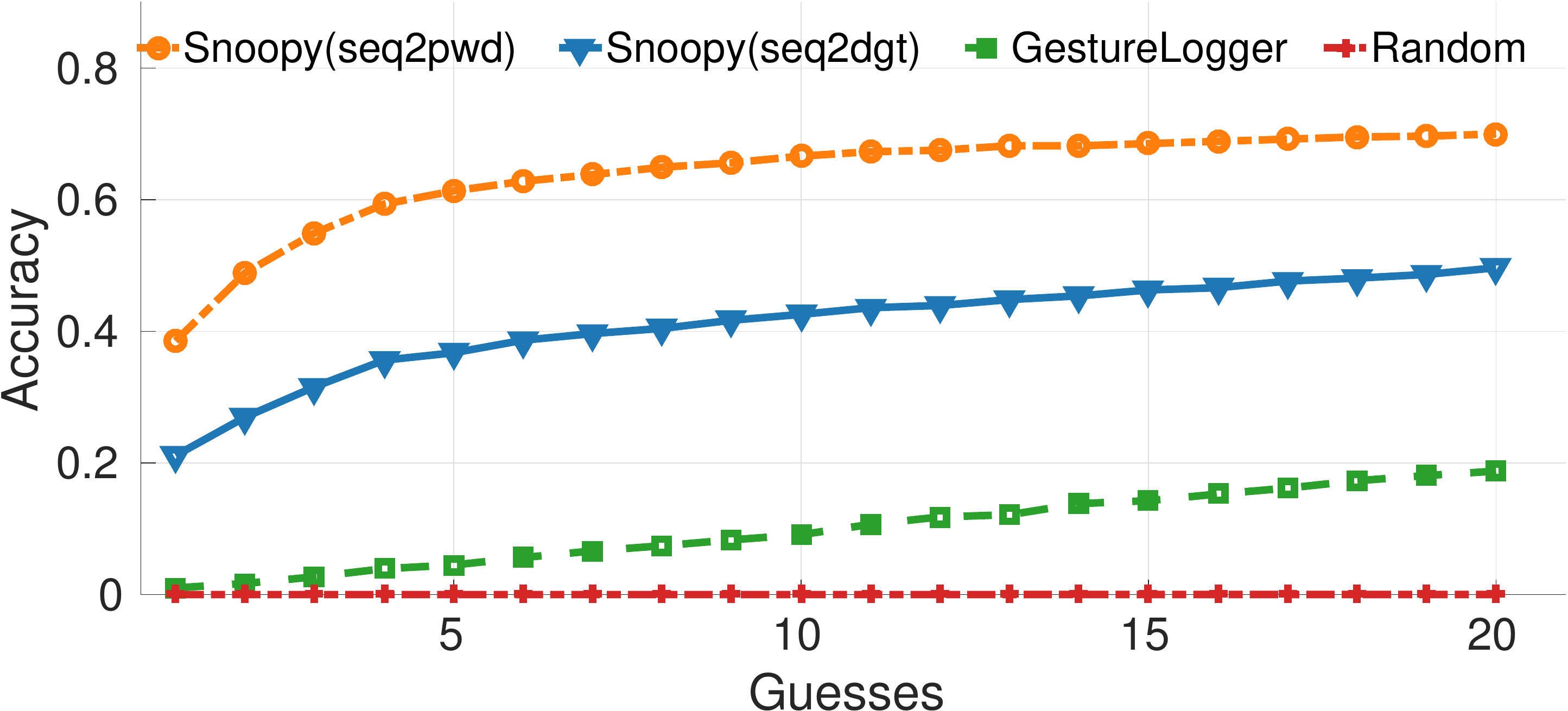}
  \caption{APL inference accuracy of competing approach and two proposed models in Snoopy.}    
     \label{fig:apl_overall_accuracy} 
  \end{minipage}%
    \hfill
  \begin{minipage}[t]{0.47\textwidth}   
    \centering   
    \includegraphics[width=\columnwidth]{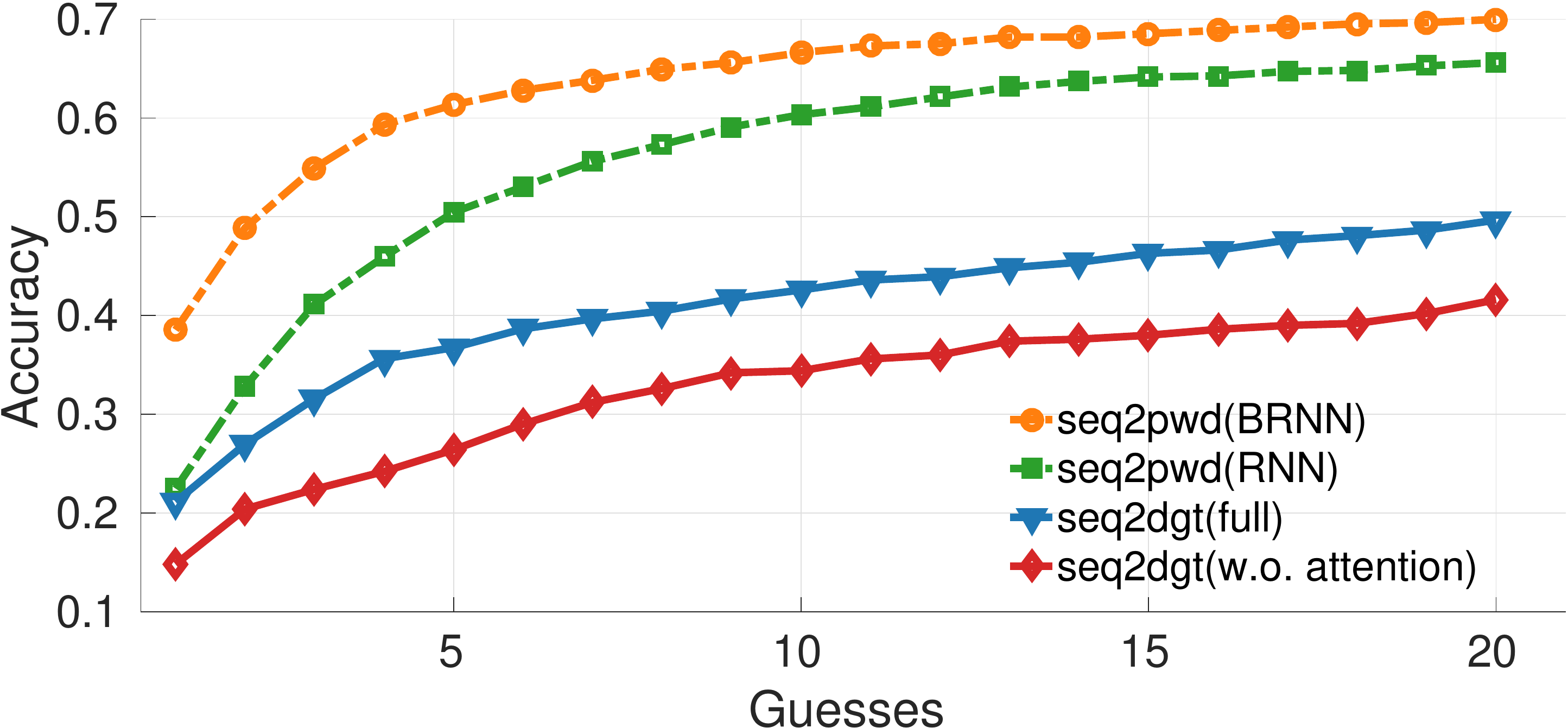} 
    \caption{Impact of network architectures on the inference accuracy.}   
    \label{fig:netwrok_architecture}
  \end{minipage} 
\end{figure*}

\subsubsection{Experiment Results}
\label{ssub:apl-results}
\hfill \break
\hongkai{\noindent \textbf{Field Test APLs vs. Constructed APL Database: }
The first experiment verifies the representativeness of the constructed APL database. We recruited an independent cohort of $308$ volunteers (115 female and 193 male, mean age 39.8 with $\sigma$ = 11.3, Mdn = 40, ranging from 18 to 63), and made sure that none of them was involved in building the APL database. Then we asked them to conduct an anonymous online survey to provide their preferred APLs. This survey also complies with data privacy policies and there is no link between the collected APLs and any individual participants. After the survey process, we obtained $308$ APL entries from all participants. We found that among the $308$ APL entries, $223$ ($72.4\%$) fall into the constructed APL database. This confirms that the constructed $P$ indeed covers a good variety of commonly used APLs, and it is possible to use $P$ to accurately infer the user entered APLs.}

\hongkai{\noindent \textbf{APL Inference Accuracy in Field Test: }
This experiment evaluates the performance of APL inference of the proposed Snoopy system in the field test. As discussed above, Snoopy uses the constructed password database $P$ and the associated motion data to train its models. To evaluate its true capability of inferring APLs in real-world scenarios, we consider the field test APLs which are independent with the APL database $P$. Concretely, we consider a similar approach as in~\cite{ye2017cracking}, and recruited another $20$ volunteers ($13$ males and $7$ females), who hadn't contributed any password or motion data, to reproduce (i.e. input) the $308$ APLs obtained from the filed test. The motion data associated with APL entries was collected using the same watch app, and on average each volunteer swiped about $120$ APLs. Eventually we obtained $2,368$ valid samples using three different types of watches (Sony SW3, Samsung Gear Live and Moto 360), and this data is the used to assess the accuracy of APL inference.

We consider the successful rate at different number of attempts~\cite{uellenbeck2013quantifying,massey1994guessing} as the metric inference accuracy, which has been widely used to quantify the threat level of a malicious app~\cite{harbach2016anatomy}. As in GestureLogger~\cite{aviv2012practicality}, we set the maximum possible number of attempts to $20$.Both proposed and competing methods take as input a motion signal sequence, and return scores for different candidate passwords. We then select the top 20 passwords, which are the most likely passwords according to the technique used. The first guess always selects the top password, the second guess the next most likely, and so on.

Fig.~\ref{fig:apl_overall_accuracy} shows inference accuracy of APLs, where we include random guess as the naive baseline. We see that both of the proposed models (seq2pwd and seq2dgt) in Snoopy consistently outperform GestureLogger, achieving up to 3-4 fold improvement in inference accuracy. In particular, if only allowed to guess once, seq2dgt model can get $21\%$ accuracy, i.e. one in five times it is able to guess the correct APL, while seq2pwd can achieve an even higher accuracy of $39\%$. We found that although seq2pwd model can only predict APLs within the constructed database $P$ ($|P|=147$), its inference accuracy is `worryingly' good: if $10$ guesses are allowed, its accuracy can be $65\%$ and increases up to to $68\%$ for $20$ guesses. Note that here the inference is performed on the field test data which is completely independent from the data used to construct $P$. This means that the APL database $P$ constructed in our experiments is very representative, and thus in practice, it is possible to infer most of the popular APLs with such a database $P$. In addition, although GestureLogger also infers APLs from $P$, its accuracy is very limited and only able to reach $19\%$ after 20 attempts (more than 3 folds lower than seq2pwd). 

On the other hand, seq2dgt is not limited to the size of database $P$, and can predict any APLs within all the $389,112$ possibilities. We see that although the search space now is $\sim2700$ times bigger, seq2dgt can still achieve decent inference accuracy: about $43\%$ after $10$ attempts and up to $50\%$ with 20 guesses. This indicates that the proposed seq2dgt model can indeed learn the underlying mechanism of user entering APLs, and make informed predictions when applicable. Note that although seq2dgt solves a much more challenging problem, i.e. no prior knowledge on popular APLs or perfect segmentation between digits, its accuracy is still way superior than the state-of-the-art GestureLogger: within $20$ guesses, seq2dgt is $250\%$ more likely to hit the correct password than GestureLogger.
}

\hongkai{\noindent \textbf{Impact of Network Architecture: } 
This experiment investigates the inference performance of Snoopy when using different deep network architectures. For seq2pwd model, we compare the inference accuracy of the proposed bi-directional RNN (B-RNN) and standard RNN. As shown in Fig.~\ref{fig:netwrok_architecture}, B-RNN is about $15\%$ superior to standard RNN at the first attempt, and is $\sim8\%$ more accurate on average within 20 attempts. This means that the temporal correlations within APLs are difficult to be captured by standard RNNs, and by allowing gradient flow from both directions, B-RNN is able to capture richer information especially in long APL sequences. On the other hand, for the seq2dgt model, we see that the proposed attention mechanism can improve about $10\%$ of inference accuracy over the standard architecture. This is because the attention mechanism helps the network to focus more on the chunks of informative sensor readings, i.e. when finger tips slide through digits, while the standard network only decodes APLs based on fixed context vectors.
}

\begin{figure*}[t]
\centering
\begin{subfigure}[b]{0.47\textwidth}\centering
\includegraphics[width=\columnwidth]{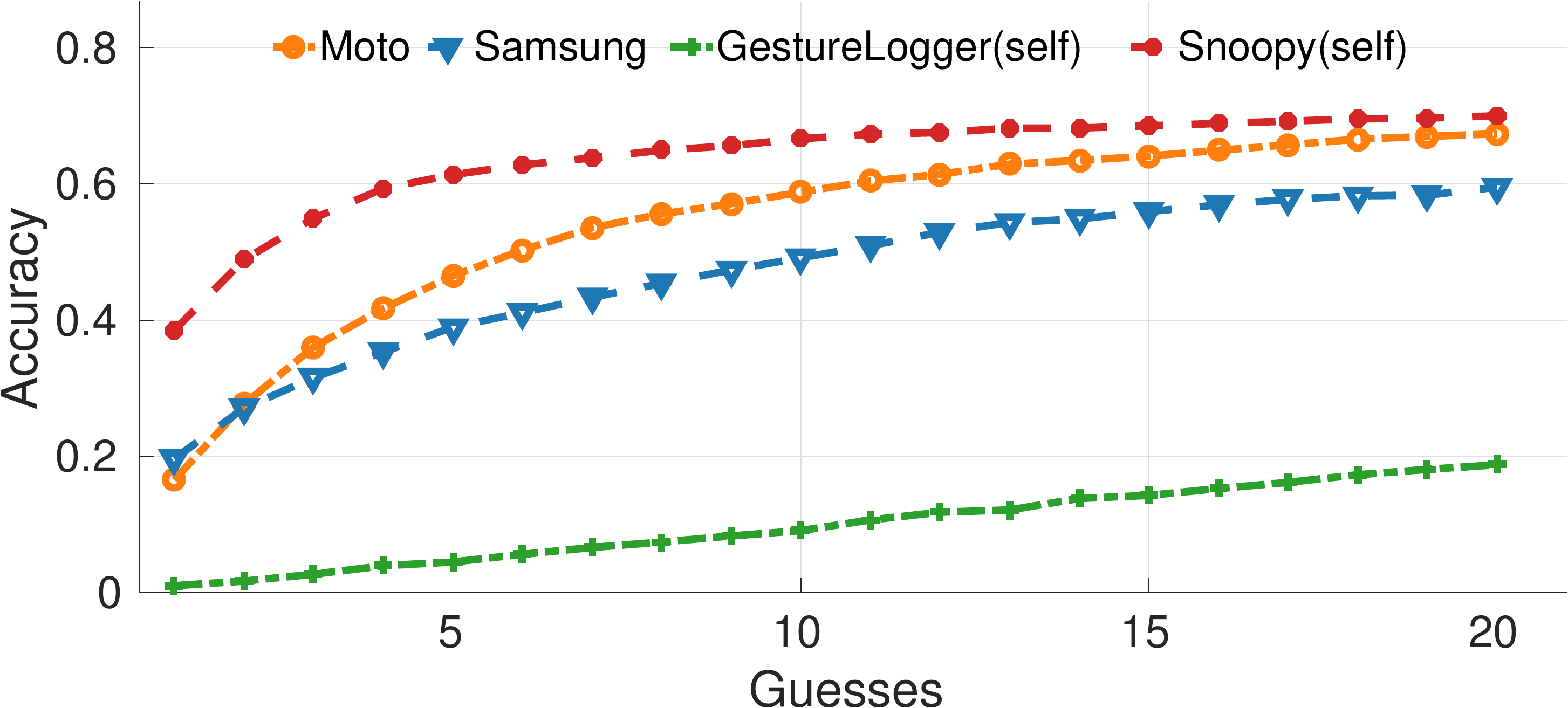}  
\end{subfigure}%
\hfill
\centering
\begin{subfigure}[b]{0.47\textwidth}\centering
\includegraphics[width=\columnwidth]{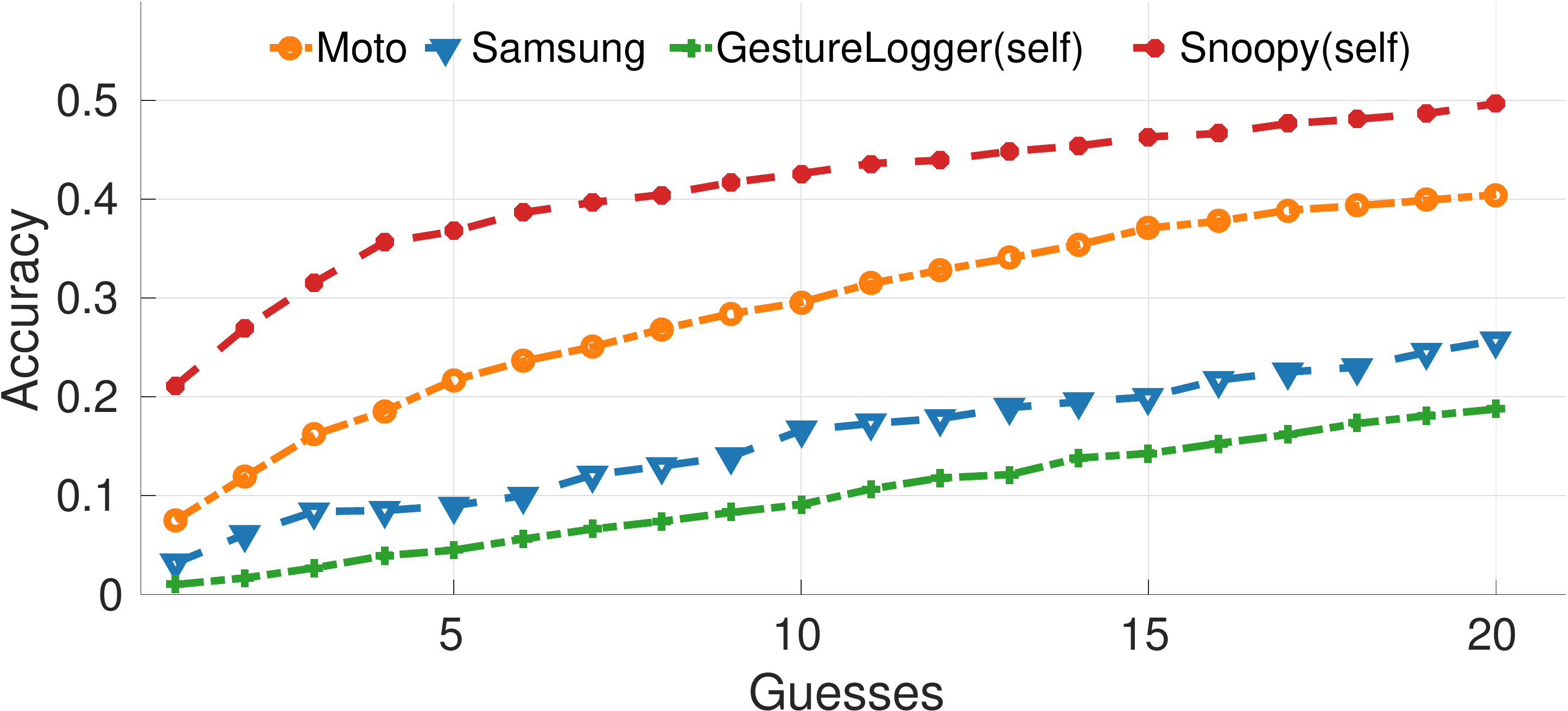} 
\end{subfigure}%
\caption{Cross-device APL inference accuracy. Left: seq2pwd model; Right: seq2dgt model.}
\label{fig:cross_device_recognition}
\end{figure*}

\hongkai{\noindent \textbf{Inference Accuracy vs. Device Heterogeneity: }
In previous experiments, although we always consider cross-user inference, i.e. our system is trained on data collected from one group of users, but tested against data generated from the others, we assume that the same model of device are used in training and testing. For instance, to infer passwords entered on a Sony SW3 watch, we assume that Snoopy can be trained with data collected on Sony SW3 watches (not necessarily the same one). In this experiment, we further push the limit, and see how Snoopy performs in the presence of device heterogeneity. This is very challenging, since the watches used for testing may have different sensors, dimensions or shapes (round vs. square) with those used for training. To demonstrate this, for APLs we train Snoopy with data from Sony SW3 watches, and test it on the other two models, Samsung Gear Live and Moto 360 Sports respectively. Fig.~\ref{fig:cross_device_recognition} shows the inference accuracy with the same device model, and across difference models. Note that here we put the performance of GestureLoger (trained and tested on the \emph{same} device model) as the baseline. As we can see, for seq2pwd, when tested on different devices, its performance drops elegantly. For Moto 360 which has round shape (the Sony watches used for training are square shaped), the performance only decreases by $\sim$13\% on average. This means heterogeneity in the shape of smartwatches won't affect password inference performance significantly. On the other hand, the performance on Samsung Gear Live (square shaped) drops about $20\%$. Note that even for this worst case, the inference accuracy of seq2pwd can still reach $\sim$50\% after 10 attempts, while the best competing approach GestureLoger is less than 10\%. On the other hand, from the right of Fig.~\ref{fig:cross_device_recognition}, we see that seq2dgt model is slightly more sensitive to device heterogeneity. On Moto watches the accuracy decreases $\sim 18\%$ while about $25\%$ on the Samsung watches. This is also expected since seq2dgt works against a massive search space ($389,112$ possibilities), where a small perturbation in sensor readings might lead to very different predictions. However even in this challenging case, the inference accuracy is still consistently higher than that of GeastureLogger, which is trained and tested on the same device models. 
}

\subsection{Performance of PIN Inference} 
\label{pin-eval}
\hongkai{In this section, we further evaluate the performance of the proposed Snoopy system in inferring PINs entered on smartwatches. }

\subsubsection{Experiment Setup} 
\label{ssub:pin-setup}
\hfill \break
\hongkai{\noindent \textbf{Training Data Collection: } 
Like the previous APL case, we first constructed a PIN database by surveying the same $112$ anonymous participants. We follow the same data collection protocol as discussed in the previous section, and obtained a database $P$ containing $79$ distinct PINs from the $112$ responses. To collect the PIN input motion data, we recruited a group of $156$ users, and ask them to input $6$ randomly selected PINs from $P$ with smartwatches (iWatches) worn on their left wrists. We use a similar data collection app, and eventually collected $23,144$ valid samples of motion data associated with the PIN input events. As in the APL case, this data is used to train the proposed Snoopy system.
}

\hongkai{\noindent \textbf{Competing Approaches: } To the best of our knowledge, there is no existing work studying PIN inference on smartwatches. Therefore, we compare Snoopy with the state-of-the-art PIN inference approaches designed for smartphones. These approaches usually adopt an element-wise inference: it firstly identifies each digit of the PINs and then concatenate the identified elements into whole passwords.
We implemented three well-known approaches: 1) \textbf{Accessory}~\cite{owusu2012accessory}: which uses a random forest classifier to identify the individual tapped digits from accelerator data; 2) \textbf{TapLogger}~\cite{xu2012taplogger} which is very similar to Accessory but uses $k-$NN classifier; and 3) \textbf{TapPrints}~\cite{miluzzo2012tapprints}, which considers both acceleration and gyroscope data, and uses an ensemble classifier (SVC, decision tree, logistic regression and random forests as base learners) to detect PIN elements. Details of the competing approaches can be found in Tab.~\ref{tab:works_on_smartphone}. Note that all of the three competing approaches require prior knowledge on the accurate segmentation of motion data, while Snoopy is able to perform end-to-end inference.
}

\begin{figure*}[t]
\centering
\begin{subfigure}[b]{0.45\textwidth}\centering
\includegraphics[width=\columnwidth]{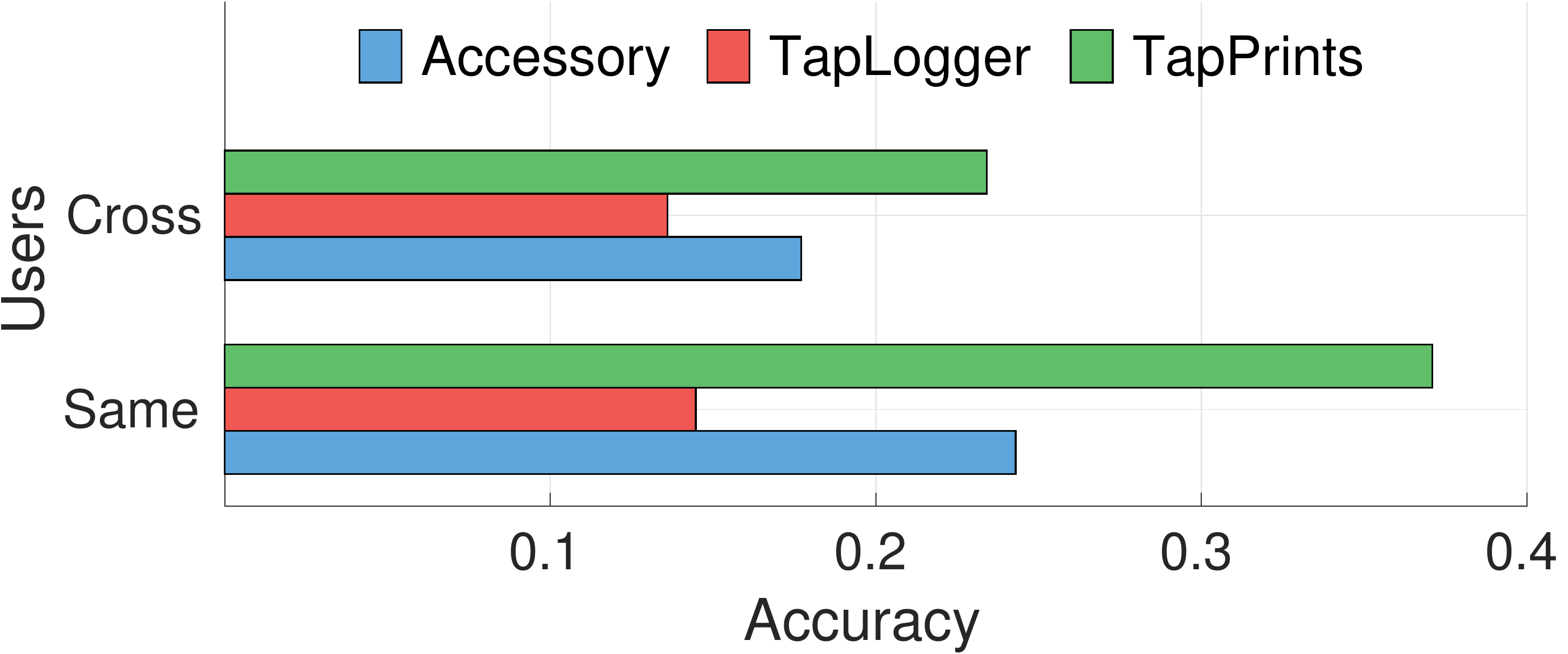}  
\end{subfigure}%
\hfill
\begin{subfigure}[b]{0.45\textwidth}\centering
\includegraphics[width=\columnwidth]{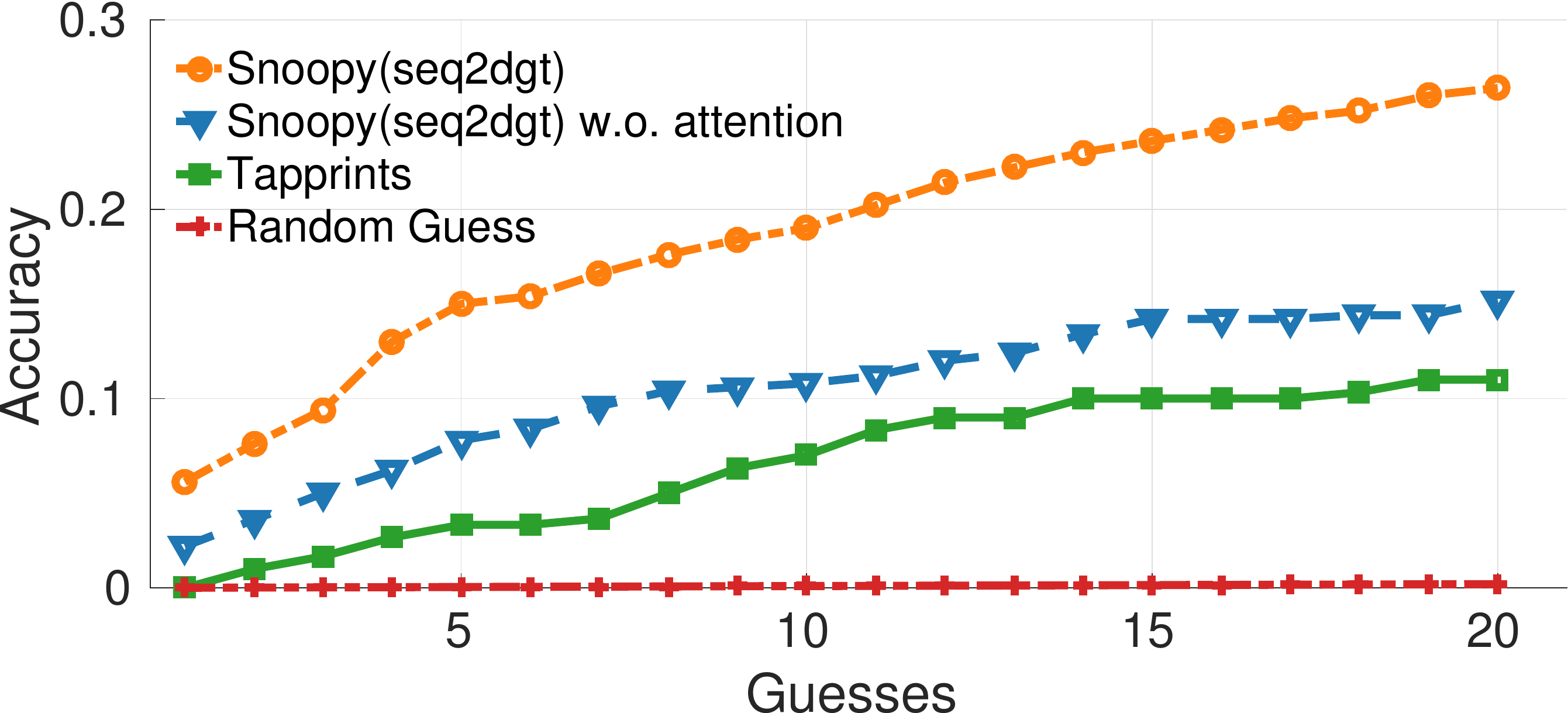} 
\end{subfigure}%
\caption{Performance of PIN inference. Left: element-wise accuracy of existing approaches; Right: inference accuracy of Snoopy and competing approaches.}
\label{fig:ios_result}
\end{figure*}

\subsubsection{Experiment Results}
\label{ssub:pin-results}
\hfill \break
\hongkai{\noindent \textbf{Field Test PINs vs. Constructed PIN Database: }
The first experiment is to evaluate to what extend can the constructed PIN database cover the commonly used PINs in the real world. We obtain from~\cite{iphone} a large online surveyed PIN dataset containing $204,508$ PINs, and consider it as the field test PIN data. As in the APL case, we rank those PINs according to their frequencies, and then select the top $642$ distinct most popular PINs to cover half of all the PIN entries ($>100k$). For those $642$ distinct PINs, we compare them with those in our PIN database $P$. We found that unlike the APL case where we observe a significant overlapping between the field test passwords and the constructed password database, here the overlapping is only about $7\%$. Unfortunately to the best of our knowledge there is no study so far that can provide thorough explanation on this. Our intuition is that people often use meaningful numbers to themselves as PINs, such as birthdays or addresses, which are quite unlikely to collide. In addition, clearly there is less constraints when tapping digits on touch screen than that of swiping APLs, and thus people may tend to choose from those easy-to-swipe APLs. Based on this observation, in the following we only consider \emph{seq2dgt} for PIN inference but not seq2pwd, since the latter can only predict PINs from the database $P$ which only covers a small percentage of commonly used PINs. 
}

\noindent \textbf{Element-wise Inference Accuracy: }
\hongkai{Before evaluating Snoopy, in this experiment, we first evaluate the performance of the existing element-wise inference approaches. We consider two types of password inference.} Firstly, the \emph{same-user inference} assumes that the algorithms would infer passwords entered by a user \emph{with} access to the ground truth password-input motion data of this particular user, e.g. they have previously ``seen'' the user entering passwords (i.e. knowing the password contents), and collected the corresponding motion data. On the other hand, the \emph{same-user inference} assumes the algorithms have to infer a user's passwords \emph{without} access to her previous labelled motion data. As shown in Fig.~\ref{fig:ios_result} (Left), the performance of element-wise approaches is very limited: the best algorithm can only achieve about 25\% accuracy when inferring a single digit for a PIN (one time guess), while the accuracy of random guess is 1/10. Even in the most favorable case where the testing objects are the same users in training, the performance only grows $\sim10\%$. A possible reason for the low segmentation accuracy is that the SNR of the motion data on smartwatches is much lower than that of smartphones, which limits the performance of prior art designed for smartphones significantly. Overall TapPrints outperforms the other two, and thus in the following experiments, we only include TapPrints in our competing approaches.

\hongkai{\noindent \textbf{PIN Inference Accuracy in Field Test: }
To evaluate the inference accuracy of the proposed Snoopy and competing approaches, we firstly collected a PIN input motion dataset based on the $642$ distinct PINs obtained from the field test. As in the previous section, we recruited an independent group of $20$ participants who hadn't contributed any data to enter those PINs on their smartwatches. The mean age of participants is 32.3 ($\sigma$ = 10.4, Mdn = 31, ranging from 18 to 53), and the data collection process is similar to that in the previous APL case. We compare the performance of the proposed seq2dgt approach in Snoopy (with and without attention mechanism) and the best competing algorithm TapPrints on this dataset, and include the naive random guess as the baseline. As shown in Fig.~\ref{fig:ios_result} (Right), both variants of Snoopy(seq2dgt) consistently outperform the TapPrints, and is able to achieve $>2\times$ accuracy improvement. In particular, with a single chance Snoopy is able to achieve $6\%$ success rate, which is much higher than random random guess ($0.01\%$). If more attempts are allowed, Snoopy can achieve up to $18\%$ success rate after $10$ guesses and $28\%$ within $20$ attempts. The performance of the competing TapPrints is much lower, and can only make to $11\%$ after $20$ attempts. In addition, we see that in this case the attention mechanism provides more performance gain (up to about $10\%$) comparing to that in the previous APL case. This is because in the case PIN inference, the motion data associated with gaps between two taps is mostly noise, which won't provide any useful information for prediction. Therefore by using the attention mechanism, Snoopy can effectively ignore those gaps by assigning dynamic weights during decoding, i.e. it would put more weights on the data segments associated with real taps.
}

%% file: section/us.tex
\section{User Study} 
\label{sec:user_study}
\begin{figure*}[t!]
  \begin{minipage}[t]{0.32\textwidth} 
  \centering   
  \includegraphics[scale=0.19]{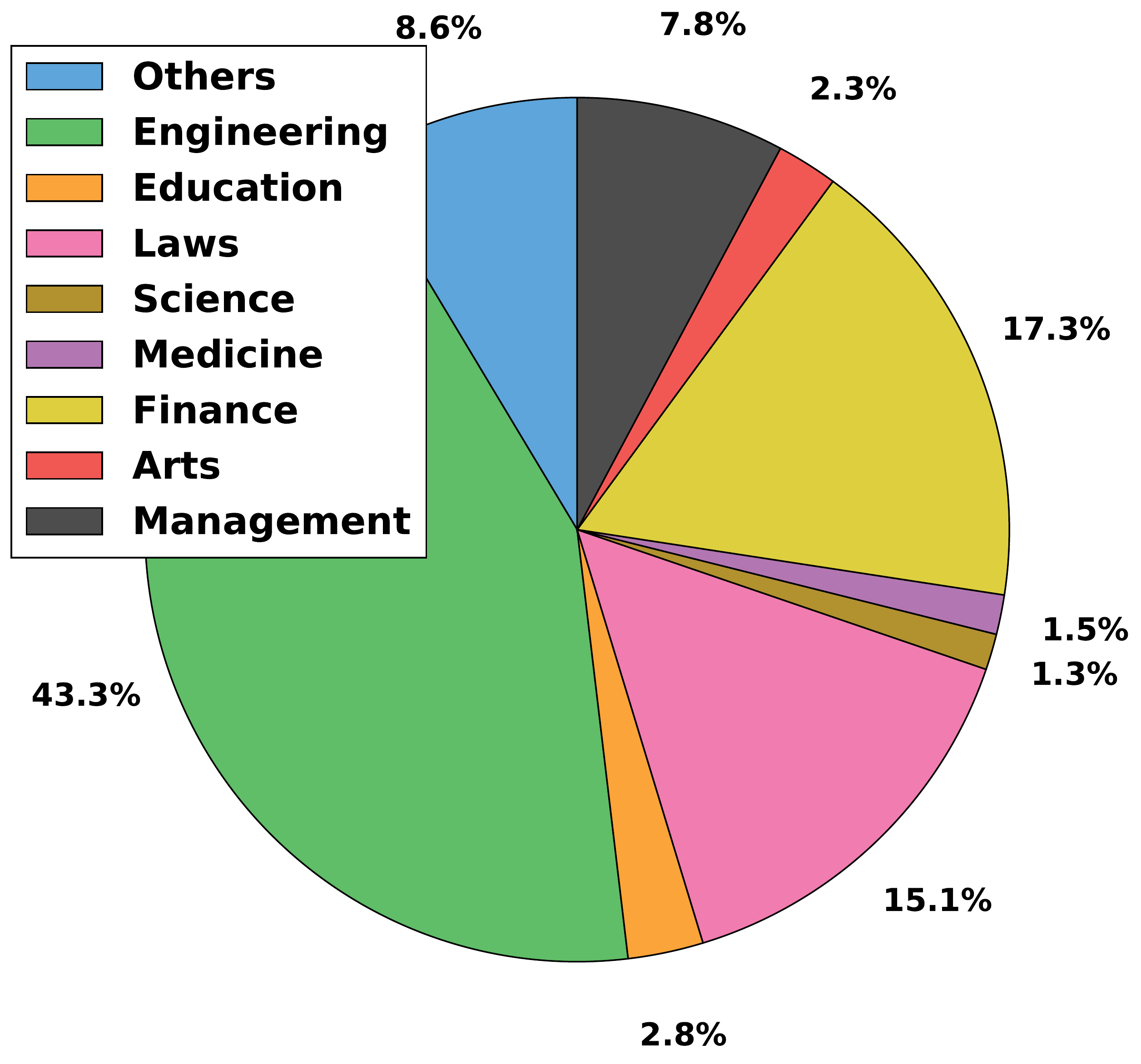}
  \caption{Background distribution.}    
     \label{fig:occupation} 
  \end{minipage}%
    \hfill
  \begin{minipage}[t]{0.32\textwidth}   
    \centering   
    \includegraphics[scale=0.19]{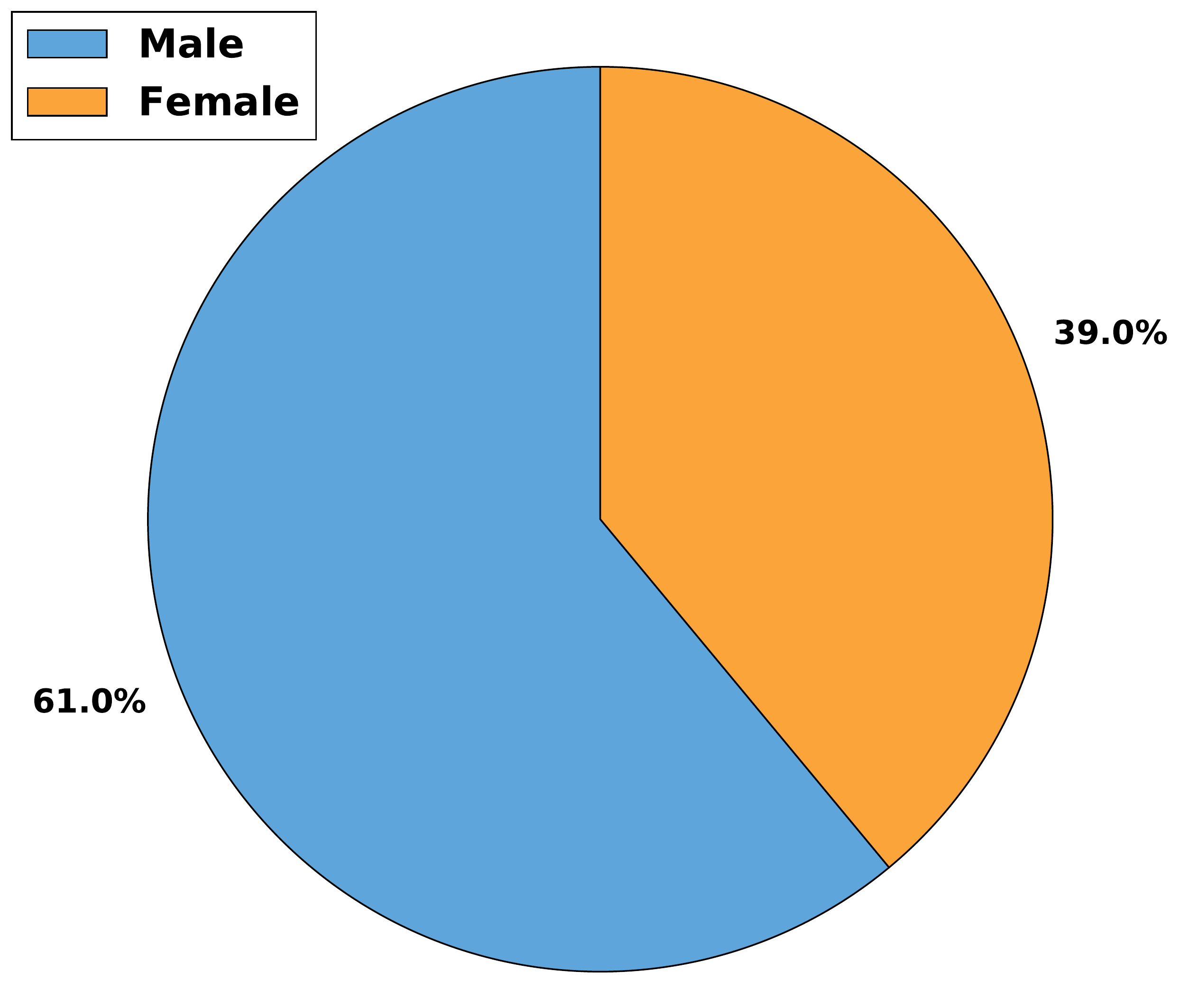} 
    \caption{Gender distribution.}   
    \label{fig:gender}
  \end{minipage} 
  \hfill
    \begin{minipage}[t]{0.32\textwidth}   
    \centering   
    \includegraphics[scale=0.19]{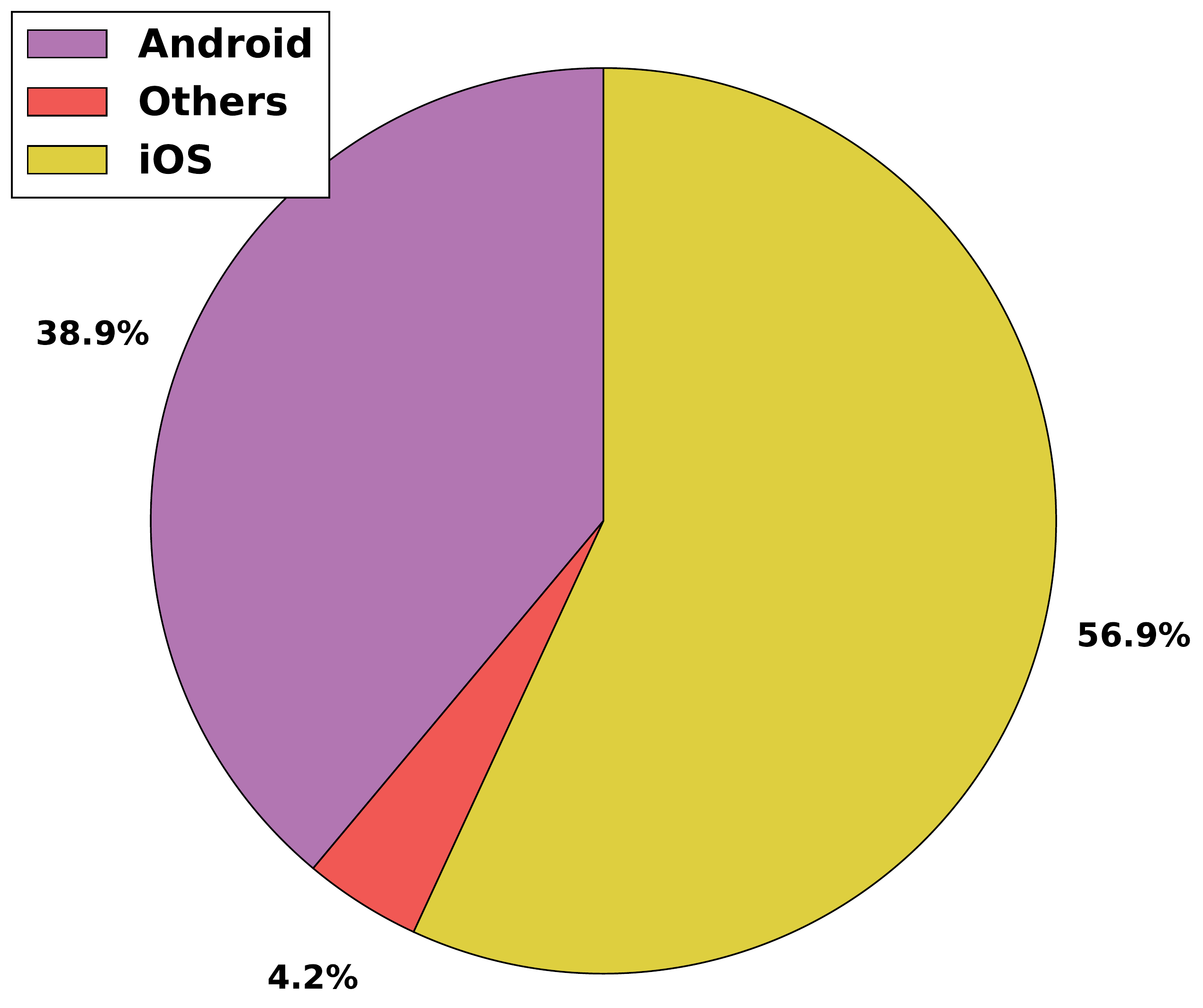} 
    \caption{Platform distribution.}   
    \label{fig:platform}
  \end{minipage}
\end{figure*}

\begin{figure*}[t!]
    \centering   
    \includegraphics[scale=.25]{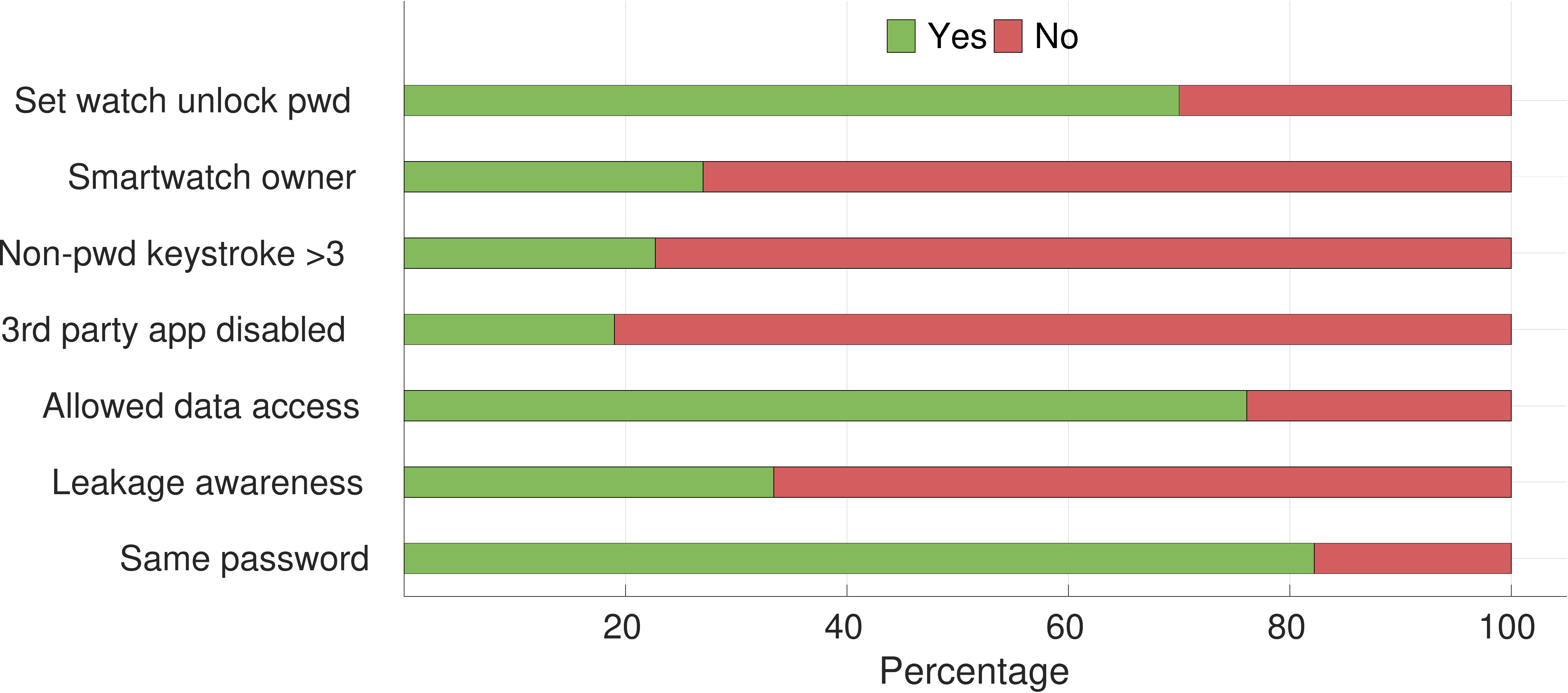} 
    \caption{Survey results. They were asked 8 questions about smartwacth usage and password settings.}   
    \label{fig:questionaire}
\end{figure*}

We complement our experimental evaluation of Snoopy with a user study, which aims to understand the users' awareness of the potential password leak on smartwatches via motion data and its consequences. We distributed questionnaires on social media in the UK and China, asking anonymous participants to provide basic demographic information such as gender, age, occupation and smartwatch platforms used. The questionnaire consists of the following yes-or-no questions:

\begin{itemize}
  \item Q1: Have you used the same passwords in different accounts/platforms, e.g., same PIN for PayPal\footnote{\url{https://www.paypal.com/gb/home}} and your device screen lock?

  \item Q2: Did you know (before this survey) your passcodes on smart devices could be leaked through motion sensors?

  \item Q3: Have you allowed or will allow third-party apps on your smart wearable devices to access your motion data, e.g., allowing WeChat+\footnote{\url{https://www.wechat.com/en/}} to record the number of steps you've walked?

  \item Q4: Have you disabled or considered disabling third-party apps monitoring your motion data when entering passwords on your smart device?

  \item Q5: Do you often type on smartwatches ($>3$ times a day), e.g., sending instant messages\footnote{\url{https://hangouts.google.com/}} or editing emails\footnote{\url{https://sparkmailapp.com/}}?
  
  \item Q6: Are you a smartwatch owner?
  
  \item Q7: Have you set up unlock passwords on your smartwatches?
  

\end{itemize}

From two periods, we received 745 anonymous responses for the first 5 questions and 301 anonymous responses for the last two questions respectively. Among them we have users from nine different occupation categories (Fig.~\ref{fig:occupation}), with slightly more male than female (61\% vs. 39\% as in Fig.~\ref{fig:gender}), and an average age of 32.3 ($\sigma$=12.1, Mdn=28, ranging from 18 to 63). This is expected since in general, young male users are more willing to try out new gadgets such as smartwatches. We also observe that Android and iOS dominate the smartwatch market. As shown in Fig.~\ref{fig:platform}, 39\% and 57\% of our participants wear Android or Apple watches, while only a tiny percentage (4\%) of them were using other platforms.

Fig~\ref{fig:questionaire} summarizes the distribution of answers. 
\hongkai{First of all, we see that about 25\% of our participants are smartwatch owners, and this number is expected to grow rapidly in the near future according to~\cite{watch2020}. Among those smartwatch owners, we find that the majority of them (73\%) would use unlock passwords for their devices. This indicates that smartwatches are truly becoming pervasive, and users tend to rely on the built-in unlock passwords (APLs or PINs) to protect their devices. Another key finding is that over $80\%$ of the participants tend to use the same password across different applications. This means that the consequences of leaking smartwatch passwords can be significant: what if the smartwatch PIN is the same PIN used for online banking or Paypal? We also observe that a large number of users ($>$60\%) did not know that the motion sensors on smart devices may leak sensitive information. In fact, 76\% of the participants would allow, or had already allowed, third-party apps to access their motion data, and less than 20\% of them would consider disabling motion sensors when entering sensitive information on their devices. This shows although motion sensor attacks have been extensively studied in academia, most users are still not aware of this. In addition, we see that unlike smartphones, users seldom perform complex interactions on smartwatch screens such as typing, since the size of them is much smaller than phones. This means in practice it is easier to detect password input events from other tapping/swiping on smartwatches, which makes this a more vulnerable attack surface. }


%% file: section/disc.tex
\section{Discussion} 
\label{sec:discussion}
\hongkai{This section discusses some important issues related to the proposed Snoopy system. In Sec.~\ref{sub:energy_accuracy_tradeoff} we show how different motion sensing strategies have knock-on effects on both energy consumption and password inference accuracy, and further highlight the benefit of the Snoopy front-end. In Sec.~\ref{sub:possible_mitigations} we discuss the limitation of the proposed approach, and provide two potential mitigations for such attacks. }

\hongkai{\subsection{Sampling More vs. Sampling Smart} }
\label{sub:energy_accuracy_tradeoff}

\hongkai{\begin{table}
\centering
\begin{tabular}{|c|c|c|c|c|}
\hline
\multirow{2}{*}{\begin{tabular}[c]{@{}c@{}}Sampling\\ Method\end{tabular}} & \multirow{2}{*}{\begin{tabular}[c]{@{}c@{}}Energy \\ Consumption\end{tabular}} & \multicolumn{2}{c|}{APL}                & PIN      \\ \cline{3-5} 
                                                                           &                                                                         & Accuracy (seq2pwd) & Accuracy (seq2dgt) & Accuracy (seq2dgt) \\ \hline
50 Hz (const)                                                                      & 2.0\% per hr                                                            & 56.1\%             & 25.7\%             & 9.1\%    \\ \hline
100 Hz (const)                                                                     & 4.4\% per hr                                                            & 61.5\%             & 34.6\%             & 14.4\%   \\ \hline
200 Hz (const)                                                                    & 6.3\% per hr                                                            & 66.3\%             & 44.5\%             & 23.4\%   \\ \hline
\textbf{Snoopy} (adaptive)                                                            & 2.1\% per  hr                                                           & 64.5\%             & 42.6\%             & 18.0\%   \\ \hline
\end{tabular}
\caption{ \hongkai{Energy consumption and inference accuracy with different motion sensing strategies (profiled on Sony SW3 watches). For the first three approaches (50Hz, 100Hz and 200Hz), we constantly sample the motion sensors and directly feed the data to the inference algorithms. The proposed Snoopy uses an adaptive sensing approach, which uses feature extraction techniques to only sample high frequency data when the users are entering passwords.} }
\label{tab:energy_accuracy_tradeoff}
\end{table}
}

\hongkai{\noindent \textbf{Energy Consumption: }
As discussed above, the front-end of Snoopy considers an adaptive motion sensing approach, which uses feature extraction techniques and SVM classifiers to detect when the user are entering passwords on their devices. Of course running feature extraction and SVM on top of motion sensing requires extra energy, however, as shown in the first two columns of Tab.~\ref{tab:energy_accuracy_tradeoff}, the overhead is not excessive: the energy consumption of Snoopy front-end is very similar to that of constantly sampling 50Hz motion data, and is only 1/3 of the 200Hz approach. Note that the adaptive sensing mechanism in Snoopy constantly tasks accelerometers at 40Hz, and triggers both accelerometer and gyroscope at 200Hz when the classifier fires. This confirms that comparing to sampling motion data at high freq, the extra energy consumption required by Snoopy front-end is cheaper.
}

\hongkai{\noindent \textbf{Password Inference Accuracy: }
We would also like to investigate the impact on password inference accuracy of different motion sensing strategies. Here we consider the inference results within 10 attempts. Obviously sampling higher frequency of motion data would help password inference, since the raw data itself contains more information. As shown in the right three columns of Tab.~\ref{tab:energy_accuracy_tradeoff}, constantly sampling 200Hz motion data gives the best password inference accuracy: 66.3\% for APL while 23.4\% for PIN. However as discussed above, the energy consumption can be prohibitively high: it would drain on average 6.3\% of battery per hour, which can easily halve the typical smartwatch lifetime \cite{sony_battery}. On the other hand, we see that by using the adaptive motion sensing strategies, Snoopy is able to achieve comparable password inference accuracy: 64.5\% for APL and 18.0\% for PIN, while keep the energy consumption much lower. We also observe that the Snoopy front-end works better for APLs than PINs: the gaps comparing to 200Hz(const) are $\sim$2\% and $\sim$5\% respectively. This is because the adaptive sampling strategy tends to have a slight delay to switch to the high sampling rate mode when the user is starting to entering passwords, and thus could lose a small chunk of motion data at the beginning. For APLs it won't affect inference much since the swiping event is continuous: one can use later data to smooth the sequence. For PINs, this is more sensitive since such data can often determine the first digit of the password and has an important impact on inference. 
}

\subsection{Possible Mitigations} 
\label{sub:possible_mitigations}
\hongkai{Like most of the other side-channel attacks, Snoopy exploits the correspondence between the leaked motion data and the contents of entered passwords. Here we propose two possible types of countermeasures.}

\hongkai{\noindent \textbf{Context-aware Motion Sensor Access: }
One promising mitigation strategy is to enable context-aware motion sensor access, as proposed in~\cite{maiti2016smartwatch}. The idea is to control the access level of motion sensors given the specific context, e.g. when users are typing on the touchscreen, it is better to limit motion sensor access, especially to those untrusted 3rd party applications. Depending on different context, one could consider control policies such as complete access blocking, reading order randomization and sampling rate reduction. However, the former two may have significant negative impact on some normal motion-dependent applications such as fitness apps, while the latter is more desirable in practice. 
}

\hongkai {\noindent \textbf{Eliminate Motion-Password Correspondence: }
The other possible countermeasure is to remove the one-to-one mapping between user generated motion data and password contents. For instance, we may insert noise to the motion data during password input, e.g. use random vibrations when the users tap or swipe, or randomly change the positions of the digits each time when users are required to authenticate, or generates different password keypads dynamically~\cite{shahzad2013secure}. Or alternatively, one can establish more types of correspondence between the users and their passwords, e.g. using the particular tapping or swiping behaviour of a user to add another layer of authentication on top of standard PINs/APLs~\cite{zheng2014you}.
}


%% file: section/related_work.tex
\section{Related Work} 
\label{sec:related_works}
\noindent \textbf{Attacking secrets via Side-channel: }
Leveraging physical sensors as a side to attack secrets has recently received lots of attention. The authors found that the MEMS gyro sensors are able to pick up low-frequency vibrations from ambient sounds. Aviv et al. demonstrated that it is possible to reconstruct a locking pattern by analyzing the oily residues left on the screen~\cite{aviv2010smudge}. This method has limited application as oily residues can be altered by other on-screen activities after pattern drawing, and also requires an attacker to have physical access to the device. Li et al. proposed a keystroke inference framework using variations in WiFi signals. They observe that keystrokes on mobile devices will lead to different hand positions and finger motion which alter the channel properties, reflected in the channel state information (CSI). A similar idea is proposed in~\cite{ali2015keystroke}, where WiFi CSI is used to infer keystrokes on a physical keyboard. However, these classes of attacks require similar environments and are highly sensitive to nearby moving objects. Vision-based attacks are also well established. Shukla et al.~\cite{shukla2014beware} used video footage captured near the victim to decode the PIN entered on the smartphone. Ye et al.~\cite{ye2017cracking} recently extend \cite{shukla2014beware} to APL and their method is robust to different lighting conditions. Though video-based side-channel attacks are very efficient in determining passwords, it is difficult for the attackers to access and locate video footage containing password input events. Compared with the above more direct attacks, eavesdropping motion sensor data is robust to environmental dynamics, is scalable, and can be achieved discreetly by a malicious app. On the hand, due to their motion tracking capability and pervasiveness, motion sensors are popular side channels for attackers. Gyrophone~\cite{michalevsky2014gyrophone} presents a new type of threat to intercept human speech by using a smartphone gyroscope. Marquardt et al. demonstrated the \emph{(Sp) iPhone}, and show that it is possible to use the accelerometer within an iPhone to recover text entered on a keyboard when the phone is placed nearby on a rigid surface~\cite{marquardt2011sp}.

\noindent \textbf{Inferring Secrets on Smart Devices via Motion Sensors: }
Researchers have attempted to infer keystrokes on smart wearables via motion sensors \cite{cai2011touchlogger,miluzzo2012tapprints,aviv2012practicality,owusu2012accessory,xu2012taplogger, Mehrnezhad2017}. The core idea behind these works is similar to the aim of Snoopy: keystrokes on device screen lead to distinct force/attitude patterns. The motion data on smart wearables can thus be used to infer entered secrets. TouchLogger \cite{cai2011touchlogger,cai2012practicality} and ACCessory \cite{owusu2012accessory} are early works, where ACCessory uses accelerometer only and TouchLogger utilizes both accelerometer and gyroscope to infer PINs. Similarly, TapLogger \cite{xu2012taplogger} refines previous techniques and uses a gyroscope to predict PIN-like secrets on smartphones. TapLogger uses a k-means clustering approach to extract the most likely classes (typically top 3). Given substantial observations of the secret (e.g. 32 PIN entry events), this is sufficient to estimate the true secret. Note that TapLogger uses manually extracted statistical features. TapPrints \cite{miluzzo2012tapprints} advances the technique and extends inference capability beyond digits to English words. The papers mentioned above require accurate digit-wise classification, which is hard to achieve with smartwatch motion data (as shown in Sec.~\ref{sec:evaluation}). In contrast, GestureLogger \cite{aviv2012practicality} infers keystrokes in an end-to-end manner rather than individually identifying each tapped digit. To this end, GestureLogger firstly designs a password database of 50 graphical passwords and 50 numerical PINs as possible passwords. It then develops a sequence classifier that infers the most likely match given the motion data sequence. However, GestureLogger uses handcrafted features for inference, which is not robust to the variability of scenarios \cite{lecun2015deep}. For example, as demonstrated in experimental results (Sec.~\ref{sec:evaluation}), the features designed for smartphones in GestureLogger did not work well in the context of smartwatches. Though redesigning new features for smartwatches is possible but the process needs domain knowledge (e.g., motion sensors). Unlike all the above, Snoopy is the only one using deep neural networks that is able to learn the best feature representations automatically through learning; its password inference framework can be easily applied to new scenarios without domain knowledge about the functioned sensors.

\chris{
While TapPrints is specifically tailored to inferring PIN passwords, we provide a uniform approach that can be used to infer both PIN and APL (swiped) passwords. Even if one focuses on PINs only, we demonstrate a ~2.5 fold improvement in accuracy compared to TapPrints. This is because TapPrints decouples the problems of segmentation and classification into two separate steps, whereas our approach handles them more robustly by tackling both tasks using the same Deep Neural Network architecture. When compared to previous work that has focused on APLs (GestureLogger), our work is fundamentally different as it can address not only APLs that exist in the training dataset, but also new previously unseen APLs. This is a significant benefit that increases the impact of the attack. As we can see in Tab.~\ref{tab:works_on_smartphone}, Snoopy is the only approach requires little domain knowledge on attackers' side, which significantly lower the bar for attack deployment. In addition, whereas all prior art has focused on smartphones, we address the problem in the context of smartwatches. This is a far more challenging scenario, due to low SNR, in which previous approaches show very low performance compared to the proposed approach. 
}


\begin{table}[!ht]
\small
\centering
\begin{tabular}{|c|c|c|c|c|c|c|c|}
\hline
\multirow{2}{*}{\textbf{Work}}                                                                        & \multirow{2}{*}{\textbf{\begin{tabular}[c]{@{}c@{}}Participants\\ Number\end{tabular}}} & \multicolumn{2}{c|}{\textbf{\begin{tabular}[c]{@{}c@{}}Domain Knowledge\\Required for Attackers\\ \end{tabular}}}                                            & \multirow{2}{*}{\textbf{\begin{tabular}[c]{@{}c@{}}Password\\ Extraction\end{tabular}}} & \multirow{2}{*}{\textbf{\begin{tabular}[c]{@{}c@{}}PWD\\ Type\end{tabular}}} & \multirow{2}{*}{\textbf{Cross-user}} & \multirow{2}{*}{\textbf{\begin{tabular}[c]{@{}c@{}}Cross-device\\ + Cross-user\end{tabular}}} \\ \cline{3-4}
                                                                                                      &                                                                                         & \textbf{\begin{tabular}[c]{@{}c@{}}Digit/swipe \\ Segmentation\end{tabular}} & \textbf{\begin{tabular}[c]{@{}c@{}}Feature\\ Engineering\end{tabular}} &                                                                                         &                                                                              &                                      &                                                                                               \\ \hline
\begin{tabular}[c]{@{}c@{}}TouchLogger\\ \cite{cai2011touchlogger,cai2012practicality}\end{tabular} & 1                                                                                       & \checkmark                                                                   & \checkmark                                                             & \xmark                                                                                  & PIN                                                                          & \xmark                               & \xmark                                                                                        \\ \hline
\begin{tabular}[c]{@{}c@{}}ACCessory\\ \cite{owusu2012accessory}\end{tabular}                       & 4                                                                                       & \checkmark                                                                   & \checkmark                                                             & \xmark                                                                                  & PIN                                                                          & \xmark                               & \xmark                                                                                        \\ \hline
\begin{tabular}[c]{@{}c@{}}TapLogger\\ \cite{xu2012taplogger}\end{tabular}                          & 3                                                                                       & \checkmark                                                                   & \checkmark                                                             & \checkmark                                                                              & PIN                                                                          & \xmark                               & \xmark                                                                                        \\ \hline
\begin{tabular}[c]{@{}c@{}}TapPrints\\ \cite{miluzzo2012tapprints}\end{tabular}                     & 10                                                                                      & \checkmark                                                                   & \checkmark                                                             & \checkmark                                                                              & PIN                                                                          & \xmark                               & \xmark                                                                                        \\ \hline
\begin{tabular}[c]{@{}c@{}}GestureLogger\\ \cite{aviv2012practicality}\end{tabular}                 & 24                                                                                      & \checkmark                                                                   & \checkmark                                                             & \xmark                                                                                  & \begin{tabular}[c]{@{}c@{}}APL\\ PIN\end{tabular}                            & \checkmark                           & \xmark                                                                                        \\ \hline
\textit{\begin{tabular}[c]{@{}c@{}}Snoopy\\ (proposed)\end{tabular}}                                  & 362                                                                                     & \xmark                                                                       & \xmark                                                                 & \checkmark                                                                              & \begin{tabular}[c]{@{}c@{}}APL\\ PIN\end{tabular}                            & \checkmark                           & \checkmark                                                                                    \\ \hline
\end{tabular}
\caption{Comparison of related works and Snoopy in password inference.}
\label{tab:works_on_smartphone}
\end{table}

\chris{
\noindent \textbf{Smartwatch Security:} 
As an increasingly ubiquitous device, the smartwatch has triggered new security issues. Wang et al. \cite{wang2015mole} and Liu et al. \cite{liu2016lasagna} pioneered this thread and have demonstrated the feasibility of inferring keystrokes on QWERTY keyboards by smartwatches. Their idea is that the keystroke-induced motions can be read from the motion sensors as long as smartwatches are worn on victims' wrists. Similar risks are also reported on the ATM machines, where victims' digital PINs are leaked through motion sensors on smartwatches. Maiti et al. \cite{maiti2016smartwatch} proposed a context-aware protection mechanism which identifies sensitive motion events (e.g., typing on the keyboard) and trigger sensor access controller accordingly. The protection mechanism has been proved to work effectively to mitigate smartwatch based side-channel attacks with least interruption for the third-party applications. 

Though Snoopy is an attack framework based on smartwatch, the goals are fundamentally different. Above works use smartwatches as a side channel to infer secrets entered on external devices, while Snoopy thrives to infer the inputs entered into the watch through it screen. }

%% file: section/conclution.tex
\section{Conclusion} 
\label{sec:conclusion}
In this paper, we studied the problem of password leaking on smartwatches via the on-board motion sensors. Although side-channel attacks based on motion data have been widely investigated on smartphones, the problem has so far been overlooked on smartwatch platforms. As a result, users are not fully aware of the risks and potential consequences. To our knowledge, this is the first work that demonstrates the feasibility of attacking passwords (PIN and Android Pattern Lock) on smartwatches using motion sensors. The proposed Snoopy system can disguise itself as a normal app (for fitness or wellbeing monitoring), and can successfully eavesdrop motion data in the background while passwords are entered. The extracted motion data is uploaded to the cloud, where Snoopy infers the contents of passwords using deep neural networks trained with crowd-sourced data. We collected large scale datasets (3 different sites, 362 users and $>$50k password entires), and compared the performance of Snoopy with state-of-the-art methods proposed for smartphones. Our findings are: a) Snoopy can accurately extract the password-positive segments of motion data (up to 98\% accuracy) in real-time on smartwatches, without consuming significant power/computational resources; b) although it is more challenging to infer passwords on smartwatches due to the much lower signal-to-noise ratio, Snoopy is able to generalize and infers passwords unseen in training c) Snoopy is robust to user and device heterogeneity. Compared with the best competing approach, it offersoffers a $\sim3$-fold improvement in the accuracy of inferring passwords. d) in practice it is vital to select the appropriate deep network architecture and regularisation parameters for Snoopy to work well. In summary, through the use of state-of-the-art deep learning, we have presented a universal technique that works for both PINs and pattern locks, without requiring any hand-engineering of features. By lowering the barrier to attackers in terms of engineering effort, the likelihood of being able to successfully compromise smart devices becomes significantly higher, simply by harvesting innocuous motion data from victims. In light of these risks, we recommend that motion sensors should be regarded as a carrying a far higher security risk than they currently are, and should have adequate countermeasures in place or require increased permissions.